\documentclass[aps,prc,preprint,amsmath,amssymb,showpacs,preprintnumbers,superscriptaddress]{revtex4-1}
\usepackage{CJK}
\usepackage{graphicx}
\usepackage{dcolumn}
\usepackage{bm}
\usepackage{color}
\usepackage{hyperref}
\usepackage{ulem}
\allowdisplaybreaks[4]

\begin{document}
\begin{CJK*}{UTF8}{}

\title{Possible bound nuclei beyond the two-neutron drip line in the $50\leqslant Z \leqslant 70$ region}
\author{C. Pan \CJKfamily{gbsn} (潘琮)}
\affiliation{State Key Laboratory of Nuclear Physics and Technology, School of Physics, Peking University, Beijing 100871, China}

\author{K. Y. Zhang \CJKfamily{gbsn} (张开元)}
\affiliation{State Key Laboratory of Nuclear Physics and Technology, School of Physics, Peking University, Beijing 100871, China}

\author{P. S. Chong }
\affiliation{Department of Physics, The University of Hong Kong, Pokfulam 999077, Hong Kong, China}

\author{C. Heo \CJKfamily{gbsn} (许赞)}
\affiliation{Department of Physics, The University of Hong Kong, Pokfulam 999077, Hong Kong, China}

\author{M. C. Ho \CJKfamily{gbsn} (何明哲)}
\affiliation{Department of Physics, The University of Hong Kong, Pokfulam 999077, Hong Kong, China}

\author{J. Lee \CJKfamily{gbsn} (李晓菁)}
\affiliation{Department of Physics, The University of Hong Kong, Pokfulam 999077, Hong Kong, China}

\author{Z. P. Li \CJKfamily{gbsn} (李志攀)}
\affiliation{School of Physical Science and Technology, Southwest University, Chongqing 400715, China}

\author{W. Sun \CJKfamily{gbsn} (孙玮)}
\affiliation{School of Physical Science and Technology, Southwest University, Chongqing 400715, China}

\author{C. K. Tam }
\affiliation{Department of Physics, The University of Hong Kong, Pokfulam 999077, Hong Kong, China}

\author{S. H. Wong \CJKfamily{bsmi} (黃首晞)}
\affiliation{Department of Physics, The University of Hong Kong, Pokfulam 999077, Hong Kong, China}

\author{R. W.-Y. Yeung }
\affiliation{Department of Physics, The University of Hong Kong, Pokfulam 999077, Hong Kong, China}

\author{T. C. Yiu \CJKfamily{gbsn} (姚道骢)}
\affiliation{Department of Physics, The University of Hong Kong, Pokfulam 999077, Hong Kong, China}

\author{S. Q. Zhang \CJKfamily{gbsn} (张双全)}
\email{sqzhang@pku.edu.cn}
\affiliation{State Key Laboratory of Nuclear Physics and Technology, School of Physics, Peking University, Beijing 100871, China}

\begin{abstract}
Possible bound nuclei beyond the two-neutron drip line in the $50\leqslant Z \leqslant 70$ region are investigated by using the deformed relativistic Hartree-Bogoliubov theory in continuum with density functional PC-PK1.
Bound nuclei beyond the drip lines of $_{56}$Ba, $_{58}$Ce, $_{62}$Sm, $_{64}$Gd and $_{66}$Dy are predicted, forming peninsulas of stability in nuclear landscape.
Near these peninsulas, several multi-neutron emitters are predicted.
The underlying mechanism of the peninsulas of stability is investigated by studying the total energy, Fermi surface, quadrupole deformation and the single-neutron spectrum in the canonical basis.
It is found that the deformation effect is crucial for forming the peninsulas of stability, and pairing correlations are also essential in specific cases.
The dependence on the deformation evolution is also discussed.
The decay rates of multi-neutron radioactivity in Ba and Sm isotopic chains are estimated by using the direct decay model.

\end{abstract}

\date{\today}


\maketitle

\end{CJK*}

\section{Introduction}

The study of nuclei far from the $\beta$-stability line, i.e., exotic nuclei, is one of the most fascinating topics in nuclear physics from both experimental \cite{Tanihata1995PPNP,Sorlin2008PPNP,Alkhazov2011IJMPE,Tanihata2013PPNP,Savran2013PPNP} and theoretical aspects \cite{Ring1996PPNP,Vretenar2005PR,Meng2006PPNP,Meng2015JPG,Meng2016book,Zhou2017PoS,Chatterjee2018PPNP}.
In exotic nuclei, especially in those near the nucleon drip lines, many interesting phenomena have been observed, including neutron and proton halos \cite{Tanihata1985PRL,Minamisono1992PRL,Schwab1995ZPA}, changes of nuclear magic numbers \cite{Ozawa2000PRL}, and pygmy resonances \cite{Adrich2005PRL}.

Another interesting exotic phenomenon is the possible existence of bound nuclei beyond the drip line \cite{Stoitsov2003PRC}.
Starting from the $\beta$-stability line and increasing the neutron number, the nuclear binding energy keeps increasing until the neutron drip line, where the binding energy begins to fall.
However, it was found in Ref.~\cite{Stoitsov2003PRC} that with a further increase of the neutron number the binding energy may increase again, which leads to reentrant stability against neutron emissions beyond the drip line and results in a complicated ``peninsula'' in nuclear landscape.
The reentrant stabilities have been predicted in several nuclear regions, for example at around $Z=60$, 70 and 100, by using the nonrelativistic Hartree-Fock-Bogoliubov (HFB) calculations in a transformed harmonic oscillator (THO) basis \cite{Stoitsov2003PRC,Erler2012Nat}, the HFB calculations in the harmonic oscillator (HO) basis together with a mapping to the five-dimensional collective Hamiltonian \cite{Delaroche2010PRC}, and the relativistic Hartree-Bogoliubov (RHB) calculations in the HO basis \cite{Afanasjev2013PLB}.
It was suggested that such phenomenon is due to the presence of shell effects at neutron closures \cite{Stoitsov2003PRC,Erler2012Nat} and the local changes of shell structure induced by deformation \cite{Afanasjev2013PLB}.
This phenomenon was also discussed based on the nonrelativistic Hartree-Fock calculations with the Bardeen-Cooper-Schrieffer (BCS) method \cite{Gridnev2015Book}.

In the above studies, however, on the one hand, nuclear stability beyond the drip line was discussed mainly based on the two-neutron separation energy and Fermi energy, while the stability against multi-neutron emission was not mentioned.
On the other hand, pairing correlations and continuum effects play crucial roles in nuclei near or beyond the drip lines.
For these nuclei, the conventional BCS approach turns out to be insufficient, whereas the Bogoliubov transformation approach provides a well-found method to treat pairing correlations \cite{Dobaczewski1984NPA}.
Although the HO basis has achieved great successes in solving deformed HFB and RHB equations \cite{Niksic2011PPNP}, it is not appropriate for the description of nuclei near the drip lines due to its incorrect asymptotic behavior \cite{Dobaczewski1996PRC,Meng1998NPA,Zhou2000CPL,Zhou2003PRC}.
In comparison with the HO basis, the THO basis provides an improved asymptotic behavior \cite{Stoitsov1998PRC_RHB,Stoitsov1998PRC_HFB}.
Furthermore, as shown in Ref.~\cite{Zhang2013PRC}, the coordinate-space calculations in large boxes are more effective when dealing with systems having small separation energy.
This continuum HFB study \cite{Zhang2013PRC} also suggested the possibility of reentrant stability.
It would be interesting to investigate the phenomenon of reentrant stability in a relativistic continuum model.

To properly consider pairing correlations and continuum effects in the relativistic framework, the relativistic continuum Hartree-Bogoliubov (RCHB) theory was developed \cite{Meng1996PRL,Meng1998NPA}, where the spherical RHB equations are solved in the coordinate space.
The RCHB theory has been successfully applied to
describe the halo in $^{11}$Li \cite{Meng1996PRL},
predict giant halos \cite{Meng1998PRL,Meng2002PRC,Zhang2002CPL},
interpret the pseudospin symmetry in exotic nuclei \cite{Meng1998PRC,Meng1999PRC},
predict new magic numbers in superheavy nuclei \cite{Zhang2005NPA}, etc.
Based on the RCHB theory, the first nuclear mass table including continuum effects was constructed and the continuum effects on the limit of nuclear landscape were explored \cite{Xia2018ADNDT}.

To simultaneously include the effects of deformation, pairing correlations and continuum, the deformed relativistic Hartree-Bogoliubov theory in continuum (DRHBc) was developed \cite{Zhou2010PRC,Li2012PRC}, in which the deformed RHB equations are solved in a Dirac Woods-Saxon basis \cite{Zhou2003PRC}.
The successful applications of the DRHBc theory include
the prediction of shape decoupling between the core and the halo in $^{42,44}$Mg \cite{Zhou2010PRC,Li2012PRC,Sun2021SciB},
the resolving of the puzzles concerning the radius and halo configuration in $^{22}$C \cite{Sun2018PLB},
the investigation of shell evolution of C isotopes and neutron halos in $^{15,19,22}$C \cite{Sun2020NPA},
the study of particles in the classically forbidden regions for Mg isotopes \cite{Zhang2019PRC},
and the interpretation of neutron halo with a small $1s_{1/2}$ component in $^{17}$B \cite{Yang2021PRL}.
Most recently, by comparing with the AME2020 data \cite{Wang2021CPC}, the predictive power of the DRHBc theory combined with the density functional PC-PK1 \cite{Zhao2010PRC} for the masses of superheavy nuclei has been shown \cite{Zhang2021PRC}.
A nuclear mass table including simultaneously the effects of deformation, pairing correlations and continuum is under construction with the DRHBc theory \cite{Zhang2020PRC}.

In this paper, taking advantage of the DRHBc theory which can properly describe deformed exotic nuclei and the successful density functional PC-PK1 \cite{Zhao2010PRC,Zhao2012PRCmass,Lu2015PRC,Agbemava2015PRC}, possible peninsulas of stability in the region from Sn to Yb ($50 \leqslant Z \leqslant 70$) will be studied in detail.
Besides the two-neutron separation energy and Fermi energy, the stability against multi-neutron emission for nuclei beyond the drip line will be examined.
In the discussed nuclear region, the neutron drip line is expected to locate between the spherical magic numbers $N=126$ and 184 \cite{Erler2012Nat,Xia2018ADNDT}, and therefore we can focus on the influences of shell closure and deformation evolution on the stability against neutron emissions beyond the drip line.
This paper is organized as follows:
In Sec.~\ref{sec:th}, a brief theoretical framework is introduced.
The numerical details are given in Sec.~\ref{sec:num}.
The results and discussion are presented in Sec.~\ref{sec:result}.
Finally, a summary is given in Sec.~\ref{sec:summary}.

\section{Theoretical framework}
\label{sec:th}

The details of the DRHBc theory can be found in Refs.~\cite{Zhou2010PRC,Li2012PRC,Zhang2020PRC}.
Here a brief formalism is presented.
In the DRHBc theory, the mean field and pairing correlations are treated self-consistently by the relativistic Hartree-Bogoliubov (RHB) equation~\cite{Kucharek1991ZPA},
\begin{equation}
	\label{eq:RHBeq}
	\begin{pmatrix}
		\hat{h}_D - \lambda_\tau & \hat{\Delta} \\ -\hat{\Delta}^* & -\hat{h}_D^* + \lambda_\tau
	\end{pmatrix}
	\begin{pmatrix} U_k \\ V_k \end{pmatrix} = E_k
	\begin{pmatrix} U_k \\ V_k \end{pmatrix},
\end{equation}
where $\lambda_\tau$ is the Fermi energy of neutron or proton $(\tau = n,p)$, $E_k$ is the quasiparticle energy, and $U_k$ and $V_k$ are the quasiparticle wave functions.
$\hat{h}_D$ is the Dirac Hamiltonian, and in the coordinate space
\begin{equation}
	h_D(\bm{r}) = \bm{\alpha}\cdot\bm{p} + V(\bm{r}) + \beta[M+S(\bm{r})],
\end{equation}
where $S(\bm{r})$ and $V(\bm{r})$ are scalar and vector potentials, respectively.
The pairing potential $\hat{\Delta}$ reads
\begin{equation}
	\Delta(\bm{r}_1,\bm{r}_2) = V^{pp}(\bm{r}_1,\bm{r}_2) \kappa(\bm{r}_1,\bm{r}_2),
\end{equation}
where $V^{pp}$ is the pairing force, and $\kappa$ is the pairing tensor \cite{Ring1980NMBP}.

For an axially deformed nucleus with spatial reflection symmetry, the potentials and densities are expanded in terms of the Legendre polynomials,
\begin{equation}
	\label{eq:Legendre}
	f(\bm{r}) = \sum_\lambda f_\lambda(r) P_\lambda(\cos\theta), \qquad \lambda = 0,2,4,\dots, \lambda_{\max}.
\end{equation}
To properly consider the continuum effect, the deformed RHB equations \eqref{eq:RHBeq} are solved in a spherical Dirac Woods-Saxon basis \cite{Zhou2003PRC}.
After self-consistently solving the RHB equations, the total energy $E_{\mathrm{tot}}$, quadrupole deformation $\beta_2$ and other expectation values can be calculated.
The canonical basis is obtained by diagonalizing the density matrix $\hat{\rho}$ \cite{Ring1980NMBP},
\begin{equation}
	\hat{\rho} |\psi_i\rangle = v_i^2 |\psi_i\rangle,
\end{equation}
where the eigenvalue $v^2_i$ is the corresponding occupation probability of $|\psi_i\rangle$.
Due to the axial and spatial reflection symmetries, the third component $m$ of the angular momentum and parity $\pi$ are good quantum numbers to characterize the canonical single-particle levels.

\section{Numerical details}
\label{sec:num}

The present DRHBc calculations are based on the point-coupling density functional PC-PK1 \cite{Zhao2010PRC}.
The PC-PK1 has turned out to be one of the best density functionals for describing nuclear properties \cite{Zhao2012PRCmass,Agbemava2015PRC,Lu2015PRC}.
The box size $R_{\mathrm{box}}=20$ fm and the mesh size $\Delta r = 0.1$ fm are taken.
For the Dirac Woods-Saxon basis, the angular momentum cutoff $J_{\max}=23/2~\hbar$ and the energy cutoff for positive-energy states $E_{\mathrm{cut}}^+ = 300$ MeV are taken, and the number of negative-energy states is taken the same as that of positive-energy states.
The Legendre expansion truncation for potentials and densities in Eq.~\eqref{eq:Legendre} is $\lambda_{\max}=6$.
The convergence for the above numerical conditions has been checked in Refs.~\cite{Pan2019IJMPE,Zhang2020PRC}.
For the pairing channel, a density-dependent zero-range force with the pairing strength $V_0 = -325 ~ \mathrm{MeV~fm^3}$ and the pairing window of 100 MeV is taken, which can nicely reproduce the odd-even mass differences for Ca and Pb isotopes \cite{Zhang2020PRC}.

\section{Results and discussion}
\label{sec:result}


To explore the possible bound nuclei beyond the drip line, in this paper we focus on the even-even nuclei near and beyond the neutron drip line in the region of $50\leqslant Z \leqslant 70$ with the state-of-the-art DRHBc theory.
With the calculated ground-state total energy $E_{\mathrm{tot}}$, or binding energy $E_B=-E_{\mathrm{tot}}$, one can obtain the two-neutron separation energy defined as $S_{2n}(Z,N) = E_B(Z,N) - E_B(Z,N-2)$.
In a given isotopic chain, the value of $S_{2n}$ generally decreases with the increase of the neutron number $N$.
Once $S_{2n}$ changes from positive to negative, the two-neutron drip line is reached, and the last nucleus with positive $S_{2n}$ is regarded as the drip-line nucleus.
Similarly, one can define the multi-neutron separation energy $S_{xn}(Z,N) = E_B(Z,N) - E_B(Z,N-x)$.
It is obvious for a bound nucleus, $S_{2n}$ and all $S_{xn}$s must be positive.
If one nucleus has a positive $S_{2n}$ but a negative $S_{xn}$, it can be regarded as a multi-neutron emitter, i.e., stable against two-neutron emission but unstable against multi-neutron emission.


\begin{figure}[htbp]
  \centering
  \includegraphics[width=1\textwidth]{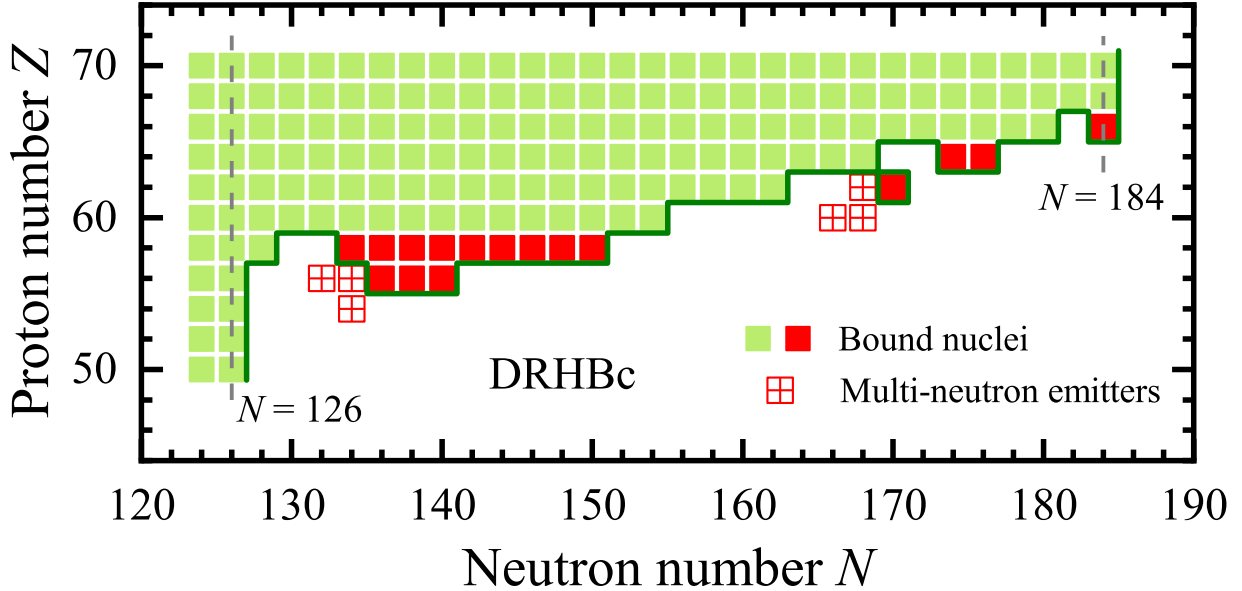}
  \caption{
  Part of nuclear landscape from Sn to Yb for even-even isotopes with the neutron number $N \geqslant 124$ from the DRHBc calculations with PC-PK1.
  Dark-green line shows the border of bound nuclei, green and red filled squares show the bound nuclei within and beyond the two-neutron drip line respectively, and red boxes with cross represent the neutron emitters stable against two-neutron emission but unstable against multi-neutron emission. }
  \label{fig:landscape}
\end{figure}

Figure \ref{fig:landscape} shows the nuclear landscape from Sn ($Z=50$) to Yb ($Z=70$) for even-even isotopes with the neutron number $N \geqslant 124$ predicted from the DRHBc calculations.
It can be found that from Sn to Ba ($Z=56$), all the two-neutron drip-line nuclei are at the traditional magic number $N=126$, reflecting a shell closure.
For Ce ($Z=58$) with two more protons, the two-neutron drip line extends to $N=128$, whereas for Nd ($Z=60$) the drip line extends dramatically to $N=154$.
It is notable that there are three Ba isotopes with $136 \leqslant N \leqslant 140$ and nine Ce isotopes with $134 \leqslant N \leqslant 150$ bound beyond the drip line.
They form an interesting peninsula of stability, with the ``northern'' boundary connected with the mainland of the nuclear landscape.
Near this peninsula, there are three multi-neutron emitters ($^{188}$Xe, $^{188,190}$Ba), which are energetically stable against two-neutron emission but unstable against multi-neutron emission.

By further increasing the proton number $Z$ from 60 to 68 (Er), as shown in Fig.~\ref{fig:landscape}, the two-neutron drip line continuously extends to the more neutron-rich region, until $N=184$, a predicted magic number \cite{Zhang2005NPA,Xia2018ADNDT}.
The two-neutron drip line of Yb is also at $N=184$.
Beyond the drip line, it is found that there are four bound nuclei, $^{232}$Sm, $^{238,240}$Gd and $^{250}$Dy, forming three small isolated regions in the nuclear landscape.
Near the bound nucleus $^{232}$Sm, three multi-neutron emitters are also found.
For Er and Yb isotopes, no bound nuclei beyond the drip line or multi-neutron emitters are found.
In the following, we will discuss the reentrant stability in detail and analyze the underlying mechanism.


\begin{figure}[htbp]
  \centering
  \includegraphics[width=0.5\textwidth]{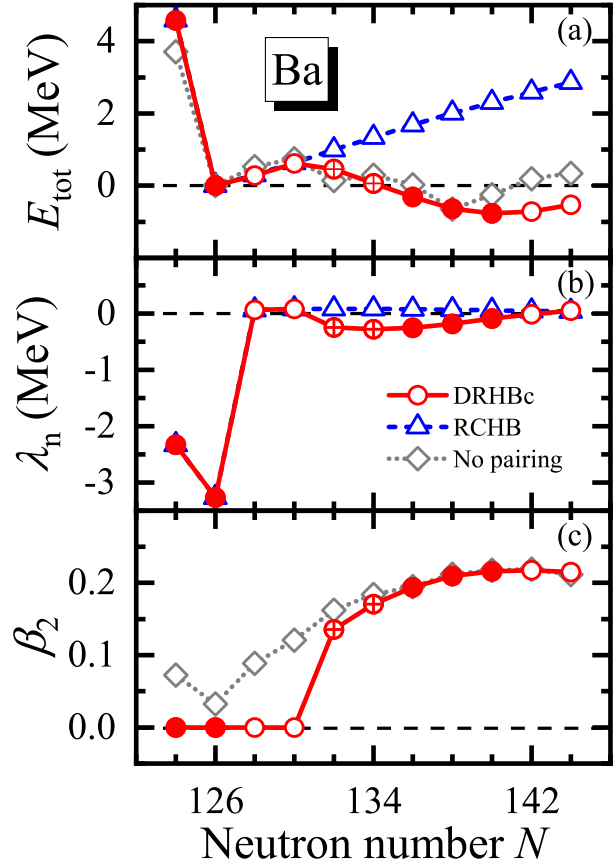}
  \caption{
  Total energy $E_{\mathrm{tot}}$ (a), neutron Fermi energy $\lambda_n$ (b) and quadrupole deformation $\beta_2$ (c) of even-even Ba isotopes as functions of the neutron number near the neutron drip line from the DRHBc calculations with PC-PK1.
  Red empty circles, red filled circles, and red boxes with cross represent respectively unbound nuclei, bound nuclei, and the neutron emitters.
  The results from the RCHB calculations (blue triangles), and the DRHBc calculations without pairing correlations (gray diamonds) are shown for comparison.
  The total energy $E_{\mathrm{tot}}$ in all calculations are shifted respectively to make them equal to 0 at the drip line.
  }
  \label{fig:Ba_bulks}
\end{figure}


Figure~\ref{fig:Ba_bulks} shows the total energy $E_{\mathrm{tot}}$, neutron Fermi energy $\lambda_n$, and quadrupole deformation $\beta_2$ for the ground states of the Ba isotopes near the drip line from the DRHBc calculations.
In Fig.~\ref{fig:Ba_bulks}(a), $E_{\mathrm{tot}}$ of the drip-line nucleus at $N=126$ is shifted to zero.
With more neutrons, the isotopes with $N=128$ and 130 loss some binding energies, i.e., $S_{2n}<0$, and thus they are unbound.
Then for $N=132$ and 134, $S_{2n}$ becomes positive, but in comparison with the drip-line nucleus at $N=126$, $S_{6n}$ or $S_{8n}$ is negative, indicating that these two nuclei are unstable against multi-neutron emission.
From $N=136$ to 140, it can be found that their $S_{2n}$ and $S_{xn}$s are all positive, showing they are energetically stable against neutron emissions and predicted to be bound nuclei beyond the drip line.
The isotopes with $N\geqslant 142$ are unbound again with $S_{2n}<0$.


The Fermi energy also carries information about the nucleon drip line, as it represents in a mean-field level the change of the total energy against the particle number~\cite{Ring1980NMBP}.
Normally, a negative Fermi energy corresponds to a positive separation energy and indicates a bound nucleus, and vice versa.
If the pairing energy vanishes, here the Fermi energy is chosen to be the energy of the last occupied single-particle state.
It is found in Fig.~\ref{fig:Ba_bulks}(b) that the neutron Fermi energy $\lambda_n$ at $N=124$ and 126 is negative, and the lower $\lambda_n$ at $N=126$ is caused by the shell closure effect, where the neutron pairing energy vanishes.
After the shell closure at $N=126$, $\lambda_n$ suddenly increases and becomes positive at $N=128$ and 130, suggesting the nuclei $^{184,186}$Ba unbound, consistent with their negative $S_{2n}$.
With more neutrons, the isotopes from $N=132$ to 140 have negative $\lambda_n$, which agrees with their positive $S_{2n}$, too.
An inconsistency between $\lambda_n$ and $S_{2n}$ is seen at $N=142$, where the very small negative $\lambda_n=-0.01$ MeV and $S_{2n}=-0.05$ MeV are obtained.
For $N=144$, $\lambda_n$ becomes positive again.

To understand the onset of bound nuclei beyond the drip line, we first analyze the evolution of $\beta_2$ in Fig.~\ref{fig:Ba_bulks}(c).
It is found that the shape of Ba isotopes keeps spherical from $N=124$ to 130, as a consequence of the strong shell closure at $N=126$.
At $N=132$ where $S_{2n}$ becomes positive and $\lambda_n$ is negative again, $\beta_2$ suddenly increases to 0.135, corresponding to an elongated spheroid.
$\beta_2$ increases gradually to 0.217 with $N=132$ to 140, which overlaps the range of the nuclei with positive $S_{2n}$ beyond the drip line.
This implies a close relationship between the bound nuclei beyond the drip line and the deformation effect.
For a further analysis, Figs.~\ref{fig:Ba_bulks}(a) and (b) also present $E_{\mathrm{tot}}$ and $\lambda_n$ obtained by using the RCHB theory where the spherical symmetry is assumed.
It is found that the RCHB drip line is at $N=126$, and no bound nucleus or multi-neutron emitter beyond the drip line is obtained.
Therefore, it is concluded that the deformation effect plays a crucial role on the emergence of the reentrant stability.


As is well known, pairing correlations are important to describe nuclear properties, especially for nuclei near the drip lines \cite{Meng2006PPNP}.
To study the role of pairing correlations, $E_{\mathrm{tot}}$ and $\beta_2$ obtained from the DRHBc calculations with pairing correlations switched off are included in Figs.~\ref{fig:Ba_bulks}(a) and (c).
It is found that the drip line also lies at $N=126$, and slight deformation appears for the isotopes nearby.
The global evolving behavior of $E_{\mathrm{tot}}$ beyond the drip line is similar to that from the full DRHBc calculations, however, with some specific differences:
$N=134$ isotope is no longer a multi-neutron emitter, and $N=136$ and 140 isotopes are not bound when pairing correlations are switched off.
Therefore, it is shown that pairing correlations can influence the range of the reentrant stability.

\begin{figure}[htbp]
  \centering
  \includegraphics[width=0.6\textwidth]{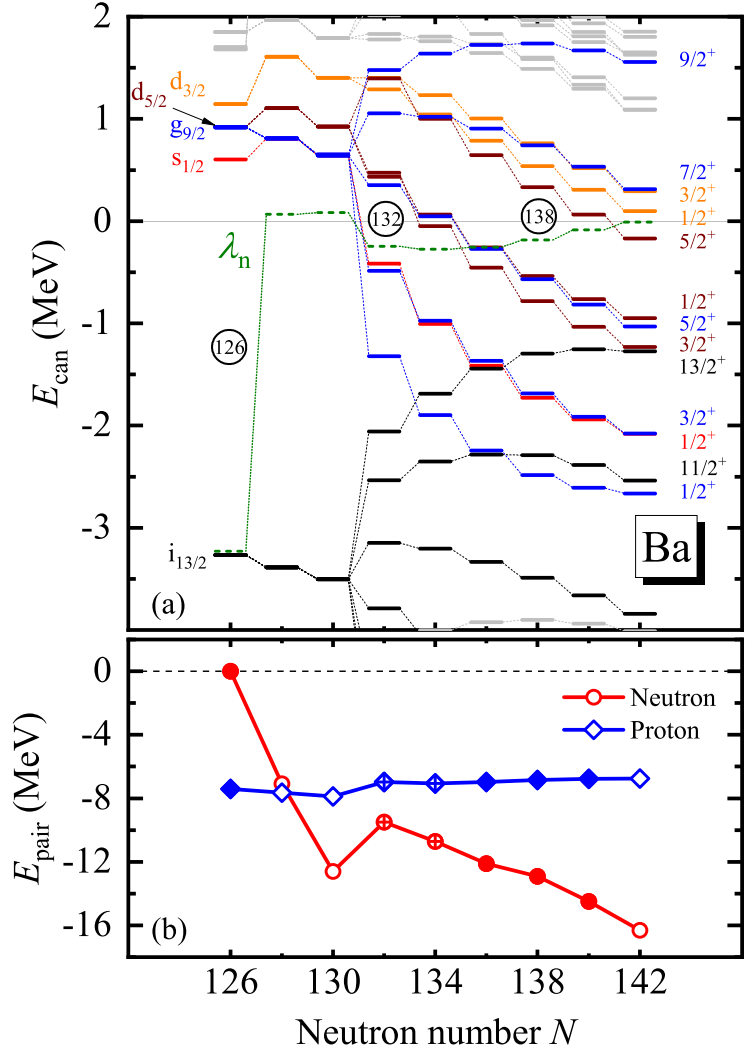}
  \caption{
  (a) Single-neutron levels around the neutron Fermi energy $\lambda_n$ (green dotted-dashed line) in the canonical basis, and (b) pairing energies
  as functions of the neutron number calculated by the DRHBc theory with PC-PK1 for even-even Ba isotopes near the neutron drip line.
  }
  \label{fig:Ba_lev}
\end{figure}

To further understand the microscopic mechanisms of the deformation and pairing correlation effects, Fig.~\ref{fig:Ba_lev}(a) shows the single-neutron energies around the neutron Fermi energy $\lambda_n$ in the canonical basis of Ba isotopes near the neutron drip line from the DRHBc calculations, and Fig.~\ref{fig:Ba_lev}(b) shows the corresponding pairing energies for proton and neutron.

As seen in Fig.~\ref{fig:Ba_lev}(a), for the spherical Ba isotopes with $N=126$ to 130, a large energy gap ($\sim 4$ MeV) above the $i_{13/2}$ orbital is clearly shown, corresponding to $N=126$ shell closure.
While the last two neutrons for $^{182}\mathrm{Ba}_{126}$ occupy the bound $i_{13/2}$ orbital with $E_{\mathrm{can}}=-3.27$ MeV, the last neutrons for $^{184}\mathrm{Ba}_{128}$ and $^{186}\mathrm{Ba}_{130}$ have to occupy the unbound orbitals ($s_{1/2}$, $g_{9/2}$) in continuum.
Correspondingly, the Fermi energy $\lambda_n$ changes from negative to positive, leaving $^{184,186}\mathrm{Ba}$ unbound and $^{182}\mathrm{Ba}$ the drip-line nucleus.
With more neutrons, the increasing occupation of the $g_{9/2}$ orbital and neighbors drives an onset of nuclear deformation, i.e., $\beta_2=0.135$ for $^{188}\mathrm{Ba}_{132}$ to $\beta_2=0.216$ for $^{196}\mathrm{Ba}_{140}$.
With the breaking of spherical symmetry, a spherical single-particle level $l_j$ splits into $(2j+1)/2$ orbitals characterized by quantum numbers $m^\pi$.
For $N=132$, the $1/2^+$ and $3/2^+$ orbitals stemming from $g_{9/2}$ as well as the $1/2^+$ from $s_{1/2}$ all lie below the continuum threshold, the occupations of which result in a negative $\lambda_n$ and a positive $S_{2n}$ of this nucleus.
By further increasing the neutron number up to $N=140$, more single-neutron levels become lower and cross the continuum threshold with the increase of deformation, which keeps $\lambda_n<0$ and $S_{2n}>0$.
In addition, one may notice in Fig.~\ref{fig:Ba_lev}(a) that the average energy of the orbitals stemming from a same $l_j$ level also moves down with the increase of the neutron number, which reflects the change of mean-field potentials with more correlated neutrons.


As seen in Fig.~\ref{fig:Ba_lev}(b), for the Ba isotopes from $N=126$ to $142$, the proton pairing energy $E_{\mathrm{pair}}^p$ keeps almost unchanged around $-8$ MeV, whereas the neutron pairing energy $E_{\mathrm{pair}}^n$ depends largely on $N$, changing from 0 to $-16$ MeV.
A sudden change of the neutron pairing energy is noticed from $N=130$ to 132, which just corresponds to the emergence of the static quadrupole deformation at $N=132$ [c.f. Fig.~\ref{fig:Ba_bulks}(c)].
The different behavior at $N=134$, 136 and 140 between the results from the DRHBc and the DRHBc without pairing, as shown in Fig.~\ref{fig:Ba_bulks}(a), can be understood as a result of the changes in the pairing energy.
When pairing correlations are switched off, $^{190}\mathrm{Ba}_{134}$ and $^{196}\mathrm{Ba}_{140}$ are unbound with $S_{2n}=-0.15$ MeV and $-0.40$ MeV, respectively, and $^{192}\mathrm{Ba}_{136}$ is a multi-neutron emitter with a positive $S_{2n}$ but negative $S_{10n}=-0.02$ MeV.
When pairing correlations are switched on, $|E_{\mathrm{pair}}^n|$ increases by 1.22 MeV from $N=132$ to 134, 1.40 MeV from 134 to 136, and 1.58 MeV from 138 to 140, which provides additional binding energies and makes $^{190}\mathrm{Ba}$ a multi-neutron emitter and $^{192,196}\mathrm{Ba}$ bound nuclei.


For the Ce isotopes, the analyses similar to Figs.~\ref{fig:Ba_bulks} and \ref{fig:Ba_lev} have been performed as well, and the same conclusion as in Ba isotopes is obtained.

\begin{figure}[htbp]
  \centering
  \includegraphics[width=1\textwidth]{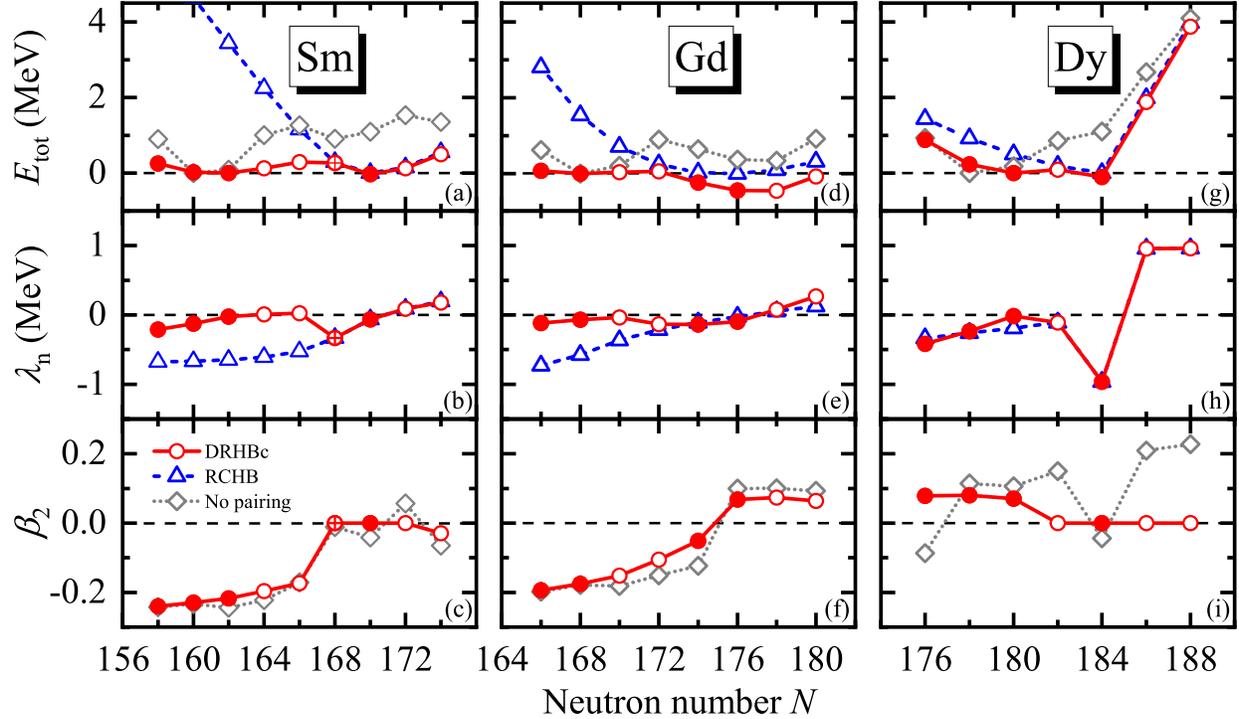}
  \caption{
  Same as Fig.~\ref{fig:Ba_bulks}, but for Sm, Gd and Dy isotopes respectively.
  }
  \label{fig:SmGdDy_bulks}
\end{figure}

Figure~\ref{fig:SmGdDy_bulks} summarizes the total energy $E_{\mathrm{tot}}$, neutron Fermi energy $\lambda_n$ and quadrupole deformation $\beta_2$ as functions of the neutron number obtained from the DRHBc calculations for Sm, Gd and Dy isotopes near the neutron drip line.
The results from the RCHB calculations and the DRHBc calculations without pairing are also included for comparison.
According to the DRHBc results of Sm isotopes in Figs.~\ref{fig:SmGdDy_bulks}(a) and \ref{fig:SmGdDy_bulks}(b), it is found that $^{224}\mathrm{Sm}_{162}$ is the drip-line nucleus, $^{232}\mathrm{Sm}_{170}$ a bound nucleus beyond the drip line, and $^{230}\mathrm{Sm}_{168}$ a multi-neutron emitter.
In Fig.~\ref{fig:SmGdDy_bulks}(c), the deformation changes suddenly from $\beta_2=-0.171$ at $N=166$ to zero at $N=168$, where $S_{2n}$ begins to be positive again.
For Gd isotopes shown in Figs.~\ref{fig:SmGdDy_bulks}(d) and \ref{fig:SmGdDy_bulks}(e), it is found that $^{232}\mathrm{Gd}_{168}$ is the drip-line nucleus, and $^{238}\mathrm{Gd}_{174}$ and $^{240}\mathrm{Gd}_{176}$ bound nuclei beyond the drip line.
In Fig.~\ref{fig:SmGdDy_bulks}(f), for $N=166$ to 174, $\beta_2$ gradually changes from $-0.19$ to $-0.05$, and for $N=176$ to 180, the shape is prolate with $\beta_2\sim 0.07$.
For Dy isotopes shown in Figs.~\ref{fig:SmGdDy_bulks}(g) and \ref{fig:SmGdDy_bulks}(h), it is found that $^{246}\mathrm{Dy}_{180}$ is the drip-line nucleus, and $^{250}\mathrm{Dy}_{184}$ a bound one beyond the drip line.
In Fig.~\ref{fig:SmGdDy_bulks}(i), for $N=176$ to 180, the shape is prolate with $\beta_2 \sim 0.1$, and for $N=182$ to 188, the shape keeps spherical.

It is notable that the shape evolution towards small deformation is shown in Figs.~\ref{fig:SmGdDy_bulks}(c, f, i) for the Sm, Gd and Dy isotopes beyond the drip line, which is in contrast with the shape evolution towards large deformation for Ba isotopes in Fig.~\ref{fig:Ba_bulks}(c).
Therefore, it is interesting to investigate how the reentrant stability depends on the shape evolution.
For this purpose, the RCHB results are also shown in Fig.~\ref{fig:SmGdDy_bulks} for comparison.
In RCHB, no bound isotope or multi-neutron emitter beyond the drip line is found.
It is interesting to note that the RCHB drip lines for Sm, Gd and Dy are more extended than the DRHBc ones, and exactly correspond to the last bound isotopes beyond the drip line predicted from DRHBc.
Indeed, it has been found in Ref.~\cite{In2021IJMPE} that the drip line is not necessarily shifted outward after the deformation effect is included, but rather depends on the evolution of deformation, for instance, the drip line ``shrinks'' accompanied by the decrease of deformation for Ne, and the opposite is seen for Ar.
Our work shows that for Sm, Gd and Dy, the decrease of deformation not only makes the drip line shrink, but also leaves some spherical (or near-spherical) isotopes bound beyond the DRHBc drip line.

The results of the DRHBc without pairing are also given in Fig.~\ref{fig:SmGdDy_bulks}, which show no bound isotope beyond the drip lines for Sm, Gd and Dy.
It means for these three isotopic chains, the deformation effect alone is insufficient to give the reentrant stability.
Here both the effects of deformation and pairing correlations are crucial to produce bound nuclei beyond the drip line, which is different from the cases of Ba and Ce.

\begin{figure}[htbp]
  \centering
  \includegraphics[width=0.6\textwidth]{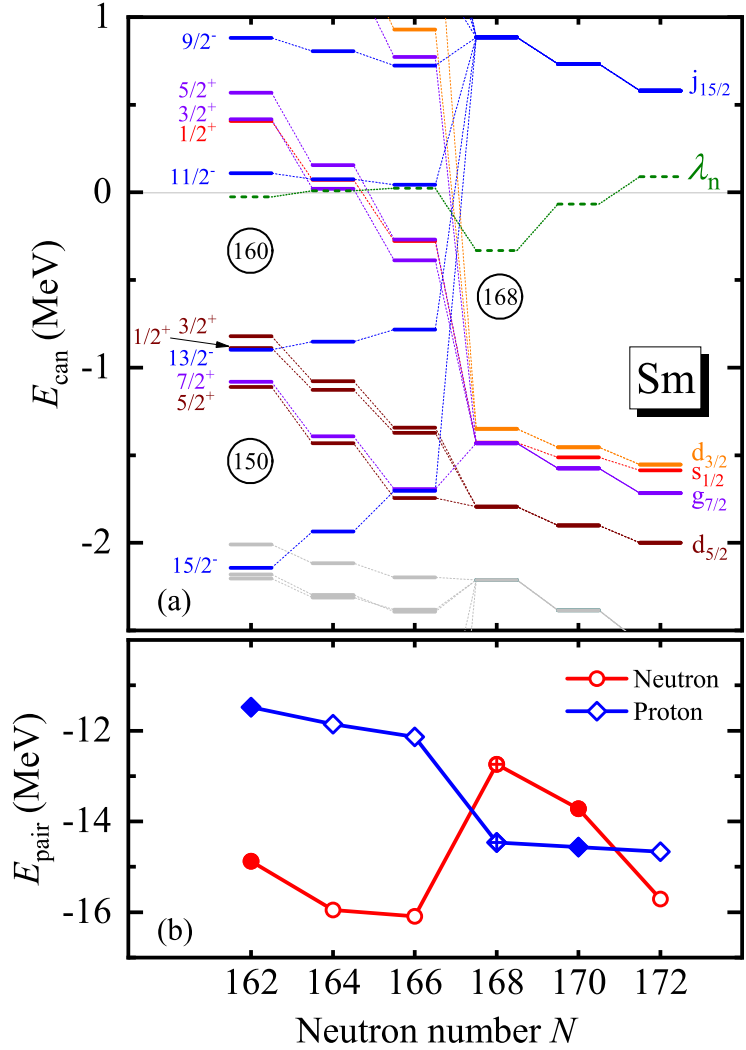}
  \caption{
  Same as Fig.~\ref{fig:Ba_lev}, but for Sm isotopes.
  }
  \label{fig:Sm_lev}
\end{figure}

More microscopically, taking Sm as an example, Fig.~\ref{fig:Sm_lev}(a) shows the evolution of the single-neutron energies around the Fermi energy $\lambda_n$ in the canonical basis with the increase of neutron number from the DRHBc calculations.
As seen in Fig.~\ref{fig:Sm_lev}(a), for the Sm isotopes with $N=162$ to 166 which have oblate deformation around $-0.2$, the neutron Fermi surface is very close to the continuum threshold and increases from a negative value at $N=162$ to positive at $N=164$ and 166.
Correspondingly, $N=162$ is the drip line, and the isotopes with $N=164$ and 166 are unbound.
By increasing the neutron number to $N=168$, the shape changes to spherical.
The splitting of single-particle levels due to deformation vanishes, and an energy gap ($\approx 2.2$ MeV) appears between the $j_{15/2}$ orbital in continuum and the bound $sdg$ orbitals, pointing to a spherical subshell at $N=168$.
The obtained $\lambda_n$ of $^{230}\mathrm{Sm}_{168}$ is $-0.33$ MeV, and the corresponding $S_{2n}$ is positive.
However, although $^{230}\mathrm{Sm}_{168}$ is stable against two-neutron emission, it is unstable against multi-neutron emission as its $E_{\mathrm{tot}}$ is higher than that of the drip-line nucleus $^{224}\mathrm{Sm}_{162}$.
For $^{232}\mathrm{Sm}_{170}$, $\lambda_n$ is still negative, and it is found to be a bound nucleus by comparing its $E_{\mathrm{tot}}$ with $^{224}\mathrm{Sm}_{162}$.
For $^{234}\mathrm{Sm}_{172}$, $\lambda_n$ becomes positive again, making it unbound.

Figure \ref{fig:Sm_lev}(b) shows the pairing energies of Sm isotopes near the drip line.
It is seen that both the neutron and proton pairing energies have a sudden change from $N=166$ to 168, which corresponds to the shape evolution from oblate to spherical.
As shown in Fig.~\ref{fig:SmGdDy_bulks}(a), when pairing correlations are switched off, $E_{\mathrm{tot}}$ of all Sm isotopes with $N>162$ are higher than that at $N=162$ by more than 0.8 MeV, and there is no bound isotope beyond the drip line.
In Fig.~\ref{fig:Sm_lev}(b), when pairing correlations are switched on, the total pairing energies $|E_{\mathrm{pair}}^n + E_{\mathrm{pair}}^p|$ of the isotopes with $164 \leqslant N \leqslant 172$ are larger than that at $N=162$, which provides additional binding energy and lowers $E_{\mathrm{tot}}$.
In particular at $N=170$, $E_{\mathrm{tot}}$ becomes lower than that at $N=162$, making $^{232}\mathrm{Sm}_{170}$ a bound isotope beyond the drip line.

In addition to the peninsulas of bound nuclei beyond the drip line, as shown in Fig. \ref{fig:landscape}, the DRHBc calculations with PC-PK1 also predict the multi-neutron emitters that are energetically stable against two-neutron emission but unstable against multi-neutron emission.
This potentially leads to the multi-neutron radioactivity, which is nowadays an interesting topic \cite{Pfutzner2012RMP,Pfutzner2013PS}.
On the experimental side, evidences for the ground-state two-neutron emitters have been reported in light nuclei $^{16}$Be \cite{Spyrou2012PRL}, $^{13}$Li \cite{Kohley2013PRC}, and $^{26}$O \cite{Kohley2013PRL}.
Meanwhile, the possibilities for the four-neutron emission in $^{7}$H, $^{18}$Be and $^{28}$O have been suggested theoretically \cite{Golovkov2004PLB,Grigorenko2011PRC}.
In $^{28}$O, a possibility of $2n$-$4n$ decay competition was also suggested \cite{Fossez2017PRC}. 
It would be illuminating to estimate the half-lives of the heavy multi-neutron emitters predicted in Ba and Sm isotopic chains and check whether their half-lives are long enough for future experimental search.

For this purpose, we adopt the direct decay model in Ref.~\cite{Grigorenko2011PRC}, which has been applied to study the two-neutron/proton and four-neutron emissions \cite{Grigorenko2011PRC,Pfutzner2012RMP,Pfutzner2013PS,Grigorenko2015PRC}.
In the direct decay model, based on the picture of independent particle, one assumes that the total decay energy $E_T$ is shared by the emitted nucleons with definite orbital angular momenta, and the interactions between the emitted nucleons are neglected.
In the four-neutron emission case, the decay width $\Gamma_{4n}$ deduced from the direct decay model reads \cite{Grigorenko2011PRC}
\begin{equation}
	\Gamma_{4n}(E_T) = \frac{E_T^3 (E_T - \sum_{i=1}^{4} E_{r_i})^2}{2\pi^3}
	\int_{0}^{1} d\epsilon_1
	\int_{0}^{1-\epsilon_1} d\epsilon_2
	\int_{0}^{1-\epsilon_1-\epsilon_2} d\epsilon_3
	\prod_{i=1}^{4} \frac{\Gamma_i(E_i)}{(E_i - E_{r_i})^2 + \Gamma_i^2(E_i)/4} .
\end{equation}
Here $E_{r_i}$ is the energy of the lowest resonance between the core and $i$-th nucleon;
$E_i = \epsilon_i E_T$ for $i\leqslant 3$ and $E_4 = (1-\sum_{i=1}^3 \epsilon_i) E_T$.
The width $\Gamma_i(E_i)$ for the two-body subsystem is given by the standard $R$-matrix expression, $\Gamma(E) = 2\gamma^2 P_l$, where $P_l$ is the penetrability.
The reduced width is $\gamma^2=\theta^2 \hbar^2/2MR^2$, with the spectroscopic factor $\theta^2$, the nucleon mass $M$, and the ``channel radius'' $R$, typically taken as $1.4 A^{1/3}$ fm \cite{Pfutzner2012RMP}.
The expression of decay width can be generalized straightforward for the six- and eight-neutron emissions.

In our estimation, the spectroscopic factor $\theta^2$ is taken as 1, and the penetrability $P_l$ for the two-body subsystem is calculated by using the WKB approximation,
\begin{equation}
	P_l = \exp\left\{ -\frac{2}{\hbar} \int_{r_1}^{r_2} \sqrt{2M [ V(r) - E_T ]} dr \right\},
\end{equation}
where $r_1$ and $r_2$ are classical turning points, and the potential $V(r) = V_N(r) + V_{\mathrm{cent}}(r)$ is the summation of the nuclear potential $V_N(r)$ obtained self-consistently from the DRHBc calculations and the centrifugal potential $V_{\mathrm{cent}}(r)$.
For simplicity only the spherical component, i.e., the $\lambda=0$ component in Eq.~\eqref{eq:Legendre}, is considered for the nuclear potential.
The resonance energies $E_{r_i}$ are taken as the difference between the calculated ground-state binding energies of the ``core'' and ``core+$n$'' systems for all emitted neutrons with different configurations.

\begin{figure}[htbp]
  \centering
  \includegraphics[width=0.8\textwidth]{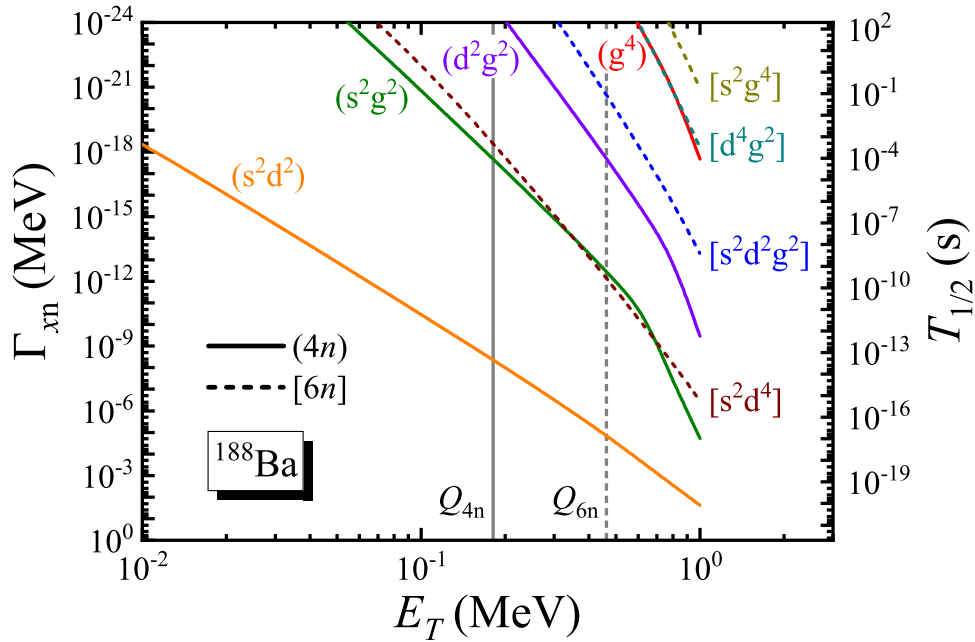}
  \caption{
  The estimated multi-neutron emission width $\Gamma_{xn}$ and the corresponding half-life $T_{1/2}$ for $^{188}$Ba as functions of the decay energy $E_T$.
  The solid and dashed lines represent the calculated results for the $4n$ and $6n$ emissions, respectively.
  The curves with different colors show the results obtained for different possible configurations of the emitted neutrons, and the parentheses () and brackets [] are used to denote $4n$ and $6n$ configurations, respectively.
  The gray vertical lines represent the calculated decay energies $Q_{4n}=0.181 ~ \mathrm{MeV}$ and $Q_{6n}=0.463 ~ \mathrm{MeV}$ for $^{188}$Ba.
   }
  \label{fig:Gamma}
\end{figure}

Taking $^{188}$Ba as an example, Fig. \ref{fig:Gamma} shows the estimated $4n$ and $6n$ emission widths $\Gamma_{xn} ~ (x=4,6)$ and the corresponding half-lives $T_{1/2}$ as functions of the decay energy $E_T$.
The decay energies of $^{188}$Ba from DRHBc calculations are $Q_{4n}=0.181$ MeV and $Q_{6n}=0.463$ MeV.
The calculated resonance energies $E_{r_i}$ are 0.621 and 0.627 MeV for the $4n$ and $6n$ emissions, respectively.
As seen in Fig. \ref{fig:Gamma}, the decay width strongly depends on the configuration of the emitted neutrons.
For $4n$ emission, the $(s^2 d^2)$ configuration is likely to provide the lower lifetime limit ($\sim 10^{-14}$ s), and the $(d^2 g^2)$ and $(g^4)$ configurations with higher orbital angular momentum components provide upper lifetime limits ($>10^2$ s).
For $6n$ emission, it shows a trend of an increasing lifetime with increasing number of emitted nucleons, and the estimated lower limit of the half-life is $\sim 10^{-10}$ s, corresponding to the $[s^2 d^4]$ configuration.
Based on the present results, $^{188}$Ba is more likely to decay via the $4n$-$2n$ sequential emission rather than the pure $6n$ emission, and the low limit of its half-life is estimated to be $\sim 10^{-14}$ s.
By means of this method, for the pure $8n$ emission of $^{190}$Ba with the extremely small decay energy $Q_{8n}=0.063$ MeV, the estimated low limit of the half-life is $\sim 10^{16}$ s, which indicates that $^{190}$Ba is more likely to decay via other modes, instead of the pure $8n$ emission.
The estimated low limits for the $4n$ and $6n$ emission half-lives of $^{230}$Sm are $\sim 10^{-12}$ and $\sim 10^{-6}$ s, respectively.
It is noted that the feasibility of an experimental search for a neutron emitter with a half-life ranging from 1 ps to sub-$\mu$s has been discussed in Ref. \cite{Grigorenko2011PRC}.
Nevertheless, we also note that the present estimations are rather simple, since the direct decay model adopted here is a very schematic model, which ignores deformation and pairing effects, and the related configuration mixing.

\section{Summary}
\label{sec:summary}

In summary, possible bound nuclei beyond the two-neutron drip line in the region from Sn to Yb are investigated by using the deformed relativistic Hartree-Bogoliubov theory in continuum with density functional PC-PK1.
The stability against multi-neutron emission is also investigated based on two-neutron and multi-neutron separation energies.
The nuclei $^{192-196}$Ba, $^{192-208}$Ce, $^{232}$Sm, $^{238,240}$Gd, and $^{250}$Dy are predicted bound beyond the drip line, forming peninsulas of stability in nuclear landscape.
It is interesting to find that near these peninsulas, some nuclei are energetically stable against two-neutron emission, but unstable against multi-neutron emission, i.e., they are multi-neutron emitters.
The decay rates of multi-neutron radioactivity predicted in Ba and Sm isotopic chains are estimated by using the direct decay model.

The underlying mechanism is investigated by studying the total energy, Fermi energy, quadrupole deformation, pairing energy and the canonical single-neutron spectra, and it is found that the deformation effect is crucial for forming the peninsulas of stability and pairing correlations are also essential in specific cases.
Interestingly, for Ba and Ce where the quadrupole deformation $\beta_2$ increases with $N$ beyond the drip line, some single-neutron levels become lower and cross the continuum threshold, which provides more binding energies and leads to the reentrant stability;
in contrast, for Sm, Gd and Dy where $\beta_2$ decreases with $N$ beyond the drip line, the drip line is less extended than the spherical case, leaving some spherical (or near-spherical) isotopes beyond the DRHBc drip line stable against specific neutron emissions.
It means that not only the increase of deformation but also the decrease of deformation can lead to the reentrant stability.
Although the locations of the reentrant stability could be model dependent \cite{Erler2012Nat,Afanasjev2013PLB}, the present discussion shows that a relativistic continuum model can hold this exotic phenomenon, and the revealed mechanism should be general.

Finally, it is interesting to note that a peninsula of stability is also found in the $100 \leqslant Z \leqslant 120$ region \cite{Zhang2021PRC,He2021CPC} according to the DRHBc theory.
It is also noted that some other effects might influence the prediction of stability beyond the drip line, such as the triaxial \cite{Lu2015PRC} and high-order deformation \cite{He2021CPC}, dynamical correlations \cite{Zhang2014FoP,Lu2015PRC,Yang2021arXiv}, charge-symmetry breaking \cite{Dong2018PRC,Dong2019NPA}, and Coulomb energy corrections \cite{Dong2019NPA,Dong2020PRC}, which deserves more efforts in the future.

\begin{acknowledgments}

Helpful discussions with members of the DRHBc Mass Table Collaboration are highly appreciated.
This work was partly supported by the National Natural Science Foundation of China (Grants No. 11935003, No. 11875075, No. 11975031, No. 12070131001, No. 11875225), the National Key R\&D Program of China (Contracts No. 2017YFE0116700 and No. 2018YFA0404400), the State Key Laboratory of Nuclear Physics and Technology, Peking University (Grant No. NPT2020ZZ01), and High-performance Computing Platform of Peking University.

\end{acknowledgments}


\begin{thebibliography}{72}%
\makeatletter
\providecommand \@ifxundefined [1]{%
 \@ifx{#1\undefined}
}%
\providecommand \@ifnum [1]{%
 \ifnum #1\expandafter \@firstoftwo
 \else \expandafter \@secondoftwo
 \fi
}%
\providecommand \@ifx [1]{%
 \ifx #1\expandafter \@firstoftwo
 \else \expandafter \@secondoftwo
 \fi
}%
\providecommand \natexlab [1]{#1}%
\providecommand \enquote  [1]{``#1''}%
\providecommand \bibnamefont  [1]{#1}%
\providecommand \bibfnamefont [1]{#1}%
\providecommand \citenamefont [1]{#1}%
\providecommand \href@noop [0]{\@secondoftwo}%
\providecommand \href [0]{\begingroup \@sanitize@url \@href}%
\providecommand \@href[1]{\@@startlink{#1}\@@href}%
\providecommand \@@href[1]{\endgroup#1\@@endlink}%
\providecommand \@sanitize@url [0]{\catcode `\\12\catcode `\$12\catcode
  `\&12\catcode `\#12\catcode `\^12\catcode `\_12\catcode `\%12\relax}%
\providecommand \@@startlink[1]{}%
\providecommand \@@endlink[0]{}%
\providecommand \url  [0]{\begingroup\@sanitize@url \@url }%
\providecommand \@url [1]{\endgroup\@href {#1}{\urlprefix }}%
\providecommand \urlprefix  [0]{URL }%
\providecommand \Eprint [0]{\href }%
\providecommand \doibase [0]{http://dx.doi.org/}%
\providecommand \selectlanguage [0]{\@gobble}%
\providecommand \bibinfo  [0]{\@secondoftwo}%
\providecommand \bibfield  [0]{\@secondoftwo}%
\providecommand \translation [1]{[#1]}%
\providecommand \BibitemOpen [0]{}%
\providecommand \bibitemStop [0]{}%
\providecommand \bibitemNoStop [0]{.\EOS\space}%
\providecommand \EOS [0]{\spacefactor3000\relax}%
\providecommand \BibitemShut  [1]{\csname bibitem#1\endcsname}%
\let\auto@bib@innerbib\@empty
\bibitem [{\citenamefont {Tanihata}(1995)}]{Tanihata1995PPNP}%
  \BibitemOpen
  \bibfield  {author} {\bibinfo {author} {\bibfnamefont {I.}~\bibnamefont
  {Tanihata}},\ }\href {\doibase 10.1016/0146-6410(95)00046-L} {\bibfield
  {journal} {\bibinfo  {journal} {Prog. Part. Nucl. Phys.}\ }\textbf {\bibinfo
  {volume} {35}},\ \bibinfo {pages} {505 } (\bibinfo {year}
  {1995})}\BibitemShut {NoStop}%
\bibitem [{\citenamefont {Sorlin}\ and\ \citenamefont
  {Porquet}(2008)}]{Sorlin2008PPNP}%
  \BibitemOpen
  \bibfield  {author} {\bibinfo {author} {\bibfnamefont {O.}~\bibnamefont
  {Sorlin}}\ and\ \bibinfo {author} {\bibfnamefont {M.-G.}\ \bibnamefont
  {Porquet}},\ }\href {\doibase 10.1016/j.ppnp.2008.05.001} {\bibfield
  {journal} {\bibinfo  {journal} {Prog. Part. Nucl. Phys.}\ }\textbf {\bibinfo
  {volume} {61}},\ \bibinfo {pages} {602 } (\bibinfo {year}
  {2008})}\BibitemShut {NoStop}%
\bibitem [{\citenamefont {Alkhazov}\ \emph {et~al.}(2011)\citenamefont
  {Alkhazov}, \citenamefont {Novikov},\ and\ \citenamefont
  {Shabelski}}]{Alkhazov2011IJMPE}%
  \BibitemOpen
  \bibfield  {author} {\bibinfo {author} {\bibfnamefont {G.~D.}\ \bibnamefont
  {Alkhazov}}, \bibinfo {author} {\bibfnamefont {I.~S.}\ \bibnamefont
  {Novikov}}, \ and\ \bibinfo {author} {\bibfnamefont {Y.~M.}\ \bibnamefont
  {Shabelski}},\ }\href {\doibase 10.1142/S0218301311018101} {\bibfield
  {journal} {\bibinfo  {journal} {Int. J. Mod. Phys. E}\ }\textbf {\bibinfo
  {volume} {20}},\ \bibinfo {pages} {583} (\bibinfo {year} {2011})}\BibitemShut
  {NoStop}%
\bibitem [{\citenamefont {Tanihata}\ \emph {et~al.}(2013)\citenamefont
  {Tanihata}, \citenamefont {Savajols},\ and\ \citenamefont
  {Kanungo}}]{Tanihata2013PPNP}%
  \BibitemOpen
  \bibfield  {author} {\bibinfo {author} {\bibfnamefont {I.}~\bibnamefont
  {Tanihata}}, \bibinfo {author} {\bibfnamefont {H.}~\bibnamefont {Savajols}},
  \ and\ \bibinfo {author} {\bibfnamefont {R.}~\bibnamefont {Kanungo}},\ }\href
  {\doibase 10.1016/j.ppnp.2012.07.001} {\bibfield  {journal} {\bibinfo
  {journal} {Prog. Part. Nucl. Phys.}\ }\textbf {\bibinfo {volume} {68}},\
  \bibinfo {pages} {215} (\bibinfo {year} {2013})}\BibitemShut {NoStop}%
\bibitem [{\citenamefont {Savran}\ \emph {et~al.}(2013)\citenamefont {Savran},
  \citenamefont {Aumann},\ and\ \citenamefont {Zilges}}]{Savran2013PPNP}%
  \BibitemOpen
  \bibfield  {author} {\bibinfo {author} {\bibfnamefont {D.}~\bibnamefont
  {Savran}}, \bibinfo {author} {\bibfnamefont {T.}~\bibnamefont {Aumann}}, \
  and\ \bibinfo {author} {\bibfnamefont {A.}~\bibnamefont {Zilges}},\ }\href
  {\doibase 10.1016/j.ppnp.2013.02.003} {\bibfield  {journal} {\bibinfo
  {journal} {Prog. Part. Nucl. Phys.}\ }\textbf {\bibinfo {volume} {70}},\
  \bibinfo {pages} {210} (\bibinfo {year} {2013})}\BibitemShut {NoStop}%
\bibitem [{\citenamefont {Ring}(1996)}]{Ring1996PPNP}%
  \BibitemOpen
  \bibfield  {author} {\bibinfo {author} {\bibfnamefont {P.}~\bibnamefont
  {Ring}},\ }\href {\doibase 10.1016/0146-6410(96)00054-3} {\bibfield
  {journal} {\bibinfo  {journal} {Prog. Part. Nucl. Phys.}\ }\textbf {\bibinfo
  {volume} {37}},\ \bibinfo {pages} {193 } (\bibinfo {year}
  {1996})}\BibitemShut {NoStop}%
\bibitem [{\citenamefont {Vretenar}\ \emph {et~al.}(2005)\citenamefont
  {Vretenar}, \citenamefont {Afanasjev}, \citenamefont {Lalazissis},\ and\
  \citenamefont {Ring}}]{Vretenar2005PR}%
  \BibitemOpen
  \bibfield  {author} {\bibinfo {author} {\bibfnamefont {D.}~\bibnamefont
  {Vretenar}}, \bibinfo {author} {\bibfnamefont {A.~V.}\ \bibnamefont
  {Afanasjev}}, \bibinfo {author} {\bibfnamefont {G.~A.}\ \bibnamefont
  {Lalazissis}}, \ and\ \bibinfo {author} {\bibfnamefont {R.}~\bibnamefont
  {Ring}},\ }\href {\doibase 10.1016/j.physrep.2004.10.001} {\bibfield
  {journal} {\bibinfo  {journal} {Phys. Rep.}\ }\textbf {\bibinfo {volume}
  {409}},\ \bibinfo {pages} {101 } (\bibinfo {year} {2005})}\BibitemShut
  {NoStop}%
\bibitem [{\citenamefont {Meng}\ \emph {et~al.}(2006)\citenamefont {Meng},
  \citenamefont {Toki}, \citenamefont {Zhou}, \citenamefont {Zhang},
  \citenamefont {Long},\ and\ \citenamefont {Geng}}]{Meng2006PPNP}%
  \BibitemOpen
  \bibfield  {author} {\bibinfo {author} {\bibfnamefont {J.}~\bibnamefont
  {Meng}}, \bibinfo {author} {\bibfnamefont {H.}~\bibnamefont {Toki}}, \bibinfo
  {author} {\bibfnamefont {S.~G.}\ \bibnamefont {Zhou}}, \bibinfo {author}
  {\bibfnamefont {S.~Q.}\ \bibnamefont {Zhang}}, \bibinfo {author}
  {\bibfnamefont {W.~H.}\ \bibnamefont {Long}}, \ and\ \bibinfo {author}
  {\bibfnamefont {L.~S.}\ \bibnamefont {Geng}},\ }\href {\doibase
  10.1016/j.ppnp.2005.06.001} {\bibfield  {journal} {\bibinfo  {journal} {Prog.
  Part. Nucl. Phys.}\ }\textbf {\bibinfo {volume} {57}},\ \bibinfo {pages} {470
  } (\bibinfo {year} {2006})}\BibitemShut {NoStop}%
\bibitem [{\citenamefont {Meng}\ and\ \citenamefont
  {Zhou}(2015)}]{Meng2015JPG}%
  \BibitemOpen
  \bibfield  {author} {\bibinfo {author} {\bibfnamefont {J.}~\bibnamefont
  {Meng}}\ and\ \bibinfo {author} {\bibfnamefont {S.-G.}\ \bibnamefont
  {Zhou}},\ }\href {\doibase 10.1088/0954-3899/42/9/093101} {\bibfield
  {journal} {\bibinfo  {journal} {J. Phys. G}\ }\textbf {\bibinfo {volume}
  {42}},\ \bibinfo {pages} {093101} (\bibinfo {year} {2015})}\BibitemShut
  {NoStop}%
\bibitem [{\citenamefont {Meng}(2016)}]{Meng2016book}%
  \BibitemOpen
  \bibinfo {editor} {\bibfnamefont {J.}~\bibnamefont {Meng}},\ ed.,\ \href
  {\doibase 10.1142/9872} {\emph {\bibinfo {title} {{Relativistic Density
  Functional for Nuclear Structure}}}}\ (\bibinfo  {publisher} {World
  Scientific},\ \bibinfo {year} {2016})\BibitemShut {NoStop}%
\bibitem [{\citenamefont {Zhou}(2017)}]{Zhou2017PoS}%
  \BibitemOpen
  \bibfield  {author} {\bibinfo {author} {\bibfnamefont {S.-G.}\ \bibnamefont
  {Zhou}},\ }\href {\doibase 10.22323/1.281.0373} {\bibfield  {journal}
  {\bibinfo  {journal} {PoS}\ }\textbf {\bibinfo {volume} {INPC2016}},\
  \bibinfo {pages} {373} (\bibinfo {year} {2017})}\BibitemShut {NoStop}%
\bibitem [{\citenamefont {Chatterjee}\ and\ \citenamefont
  {Shyam}(2018)}]{Chatterjee2018PPNP}%
  \BibitemOpen
  \bibfield  {author} {\bibinfo {author} {\bibfnamefont {R.}~\bibnamefont
  {Chatterjee}}\ and\ \bibinfo {author} {\bibfnamefont {R.}~\bibnamefont
  {Shyam}},\ }\href {\doibase 10.1016/j.ppnp.2018.06.001} {\bibfield  {journal}
  {\bibinfo  {journal} {Prog. Part. Nucl. Phys.}\ }\textbf {\bibinfo {volume}
  {103}},\ \bibinfo {pages} {67 } (\bibinfo {year} {2018})}\BibitemShut
  {NoStop}%
\bibitem [{\citenamefont {Tanihata}\ \emph {et~al.}(1985)\citenamefont
  {Tanihata}, \citenamefont {Hamagaki}, \citenamefont {Hashimoto},
  \citenamefont {Shida}, \citenamefont {Yoshikawa}, \citenamefont {Sugimoto},
  \citenamefont {Yamakawa}, \citenamefont {Kobayashi},\ and\ \citenamefont
  {Takahashi}}]{Tanihata1985PRL}%
  \BibitemOpen
  \bibfield  {author} {\bibinfo {author} {\bibfnamefont {I.}~\bibnamefont
  {Tanihata}}, \bibinfo {author} {\bibfnamefont {H.}~\bibnamefont {Hamagaki}},
  \bibinfo {author} {\bibfnamefont {O.}~\bibnamefont {Hashimoto}}, \bibinfo
  {author} {\bibfnamefont {Y.}~\bibnamefont {Shida}}, \bibinfo {author}
  {\bibfnamefont {N.}~\bibnamefont {Yoshikawa}}, \bibinfo {author}
  {\bibfnamefont {K.}~\bibnamefont {Sugimoto}}, \bibinfo {author}
  {\bibfnamefont {O.}~\bibnamefont {Yamakawa}}, \bibinfo {author}
  {\bibfnamefont {T.}~\bibnamefont {Kobayashi}}, \ and\ \bibinfo {author}
  {\bibfnamefont {N.}~\bibnamefont {Takahashi}},\ }\href {\doibase
  10.1103/PhysRevLett.55.2676} {\bibfield  {journal} {\bibinfo  {journal}
  {Phys. Rev. Lett.}\ }\textbf {\bibinfo {volume} {55}},\ \bibinfo {pages}
  {2676} (\bibinfo {year} {1985})}\BibitemShut {NoStop}%
\bibitem [{\citenamefont {Minamisono}\ \emph {et~al.}(1992)\citenamefont
  {Minamisono}, \citenamefont {Ohtsubo}, \citenamefont {Minami}, \citenamefont
  {Fukuda}, \citenamefont {Kitagawa}, \citenamefont {Fukuda}, \citenamefont
  {Matsuta}, \citenamefont {Nojiri}, \citenamefont {Takeda}, \citenamefont
  {Sagawa},\ and\ \citenamefont {Kitagawa}}]{Minamisono1992PRL}%
  \BibitemOpen
  \bibfield  {author} {\bibinfo {author} {\bibfnamefont {T.}~\bibnamefont
  {Minamisono}}, \bibinfo {author} {\bibfnamefont {T.}~\bibnamefont {Ohtsubo}},
  \bibinfo {author} {\bibfnamefont {I.}~\bibnamefont {Minami}}, \bibinfo
  {author} {\bibfnamefont {S.}~\bibnamefont {Fukuda}}, \bibinfo {author}
  {\bibfnamefont {A.}~\bibnamefont {Kitagawa}}, \bibinfo {author}
  {\bibfnamefont {M.}~\bibnamefont {Fukuda}}, \bibinfo {author} {\bibfnamefont
  {K.}~\bibnamefont {Matsuta}}, \bibinfo {author} {\bibfnamefont
  {Y.}~\bibnamefont {Nojiri}}, \bibinfo {author} {\bibfnamefont
  {S.}~\bibnamefont {Takeda}}, \bibinfo {author} {\bibfnamefont
  {H.}~\bibnamefont {Sagawa}}, \ and\ \bibinfo {author} {\bibfnamefont
  {H.}~\bibnamefont {Kitagawa}},\ }\href {\doibase 10.1103/PhysRevLett.69.2058}
  {\bibfield  {journal} {\bibinfo  {journal} {Phys. Rev. Lett.}\ }\textbf
  {\bibinfo {volume} {69}},\ \bibinfo {pages} {2058} (\bibinfo {year}
  {1992})}\BibitemShut {NoStop}%
\bibitem [{\citenamefont {Schwab}\ \emph {et~al.}(1995)\citenamefont {Schwab},
  \citenamefont {Geissel}, \citenamefont {Behr}, \citenamefont {Br\"unle},
  \citenamefont {Burkard}, \citenamefont {Irnich}, \citenamefont {Kraus},
  \citenamefont {M\"unzenberg}, \citenamefont {Nickel}, \citenamefont
  {Scheidenberger}, \citenamefont {Suzuki},\ and\ \citenamefont
  {Voss}}]{Schwab1995ZPA}%
  \BibitemOpen
  \bibfield  {author} {\bibinfo {author} {\bibfnamefont {W.}~\bibnamefont
  {Schwab}}, \bibinfo {author} {\bibfnamefont {H.}~\bibnamefont {Geissel}},
  \bibinfo {author} {\bibfnamefont {K.-H.}\ \bibnamefont {Behr}}, \bibinfo
  {author} {\bibfnamefont {A.}~\bibnamefont {Br\"unle}}, \bibinfo {author}
  {\bibfnamefont {K.}~\bibnamefont {Burkard}}, \bibinfo {author} {\bibfnamefont
  {H.}~\bibnamefont {Irnich}}, \bibinfo {author} {\bibfnamefont
  {G.}~\bibnamefont {Kraus}}, \bibinfo {author} {\bibfnamefont
  {G.}~\bibnamefont {M\"unzenberg}}, \bibinfo {author} {\bibfnamefont
  {F.}~\bibnamefont {Nickel}}, \bibinfo {author} {\bibfnamefont
  {C.}~\bibnamefont {Scheidenberger}}, \bibinfo {author} {\bibfnamefont
  {T.}~\bibnamefont {Suzuki}}, \ and\ \bibinfo {author} {\bibfnamefont
  {B.}~\bibnamefont {Voss}},\ }\href {\doibase 10.1007/BF01291183} {\bibfield
  {journal} {\bibinfo  {journal} {Z. Phys. A}\ }\textbf {\bibinfo {volume}
  {350}},\ \bibinfo {pages} {283} (\bibinfo {year} {1995})}\BibitemShut
  {NoStop}%
\bibitem [{\citenamefont {Ozawa}\ \emph {et~al.}(2000)\citenamefont {Ozawa},
  \citenamefont {Kobayashi}, \citenamefont {Suzuki}, \citenamefont {Yoshida},\
  and\ \citenamefont {Tanihata}}]{Ozawa2000PRL}%
  \BibitemOpen
  \bibfield  {author} {\bibinfo {author} {\bibfnamefont {A.}~\bibnamefont
  {Ozawa}}, \bibinfo {author} {\bibfnamefont {T.}~\bibnamefont {Kobayashi}},
  \bibinfo {author} {\bibfnamefont {T.}~\bibnamefont {Suzuki}}, \bibinfo
  {author} {\bibfnamefont {K.}~\bibnamefont {Yoshida}}, \ and\ \bibinfo
  {author} {\bibfnamefont {I.}~\bibnamefont {Tanihata}},\ }\href {\doibase
  10.1103/PhysRevLett.84.5493} {\bibfield  {journal} {\bibinfo  {journal}
  {Phys. Rev. Lett.}\ }\textbf {\bibinfo {volume} {84}},\ \bibinfo {pages}
  {5493} (\bibinfo {year} {2000})}\BibitemShut {NoStop}%
\bibitem [{\citenamefont {Adrich}\ \emph {et~al.}(2005)\citenamefont {Adrich},
  \citenamefont {Klimkiewicz}, \citenamefont {Fallot}, \citenamefont
  {Boretzky}, \citenamefont {Aumann}, \citenamefont {Cortina-Gil},
  \citenamefont {Pramanik}, \citenamefont {Elze}, \citenamefont {Emling},
  \citenamefont {Geissel}, \citenamefont {Hellstr\"om}, \citenamefont {Jones},
  \citenamefont {Kratz}, \citenamefont {Kulessa}, \citenamefont {Leifels},
  \citenamefont {Nociforo}, \citenamefont {Palit}, \citenamefont {Simon},
  \citenamefont {Sur\'owka}, \citenamefont {S\"ummerer},\ and\ \citenamefont
  {Walu\ifmmode~\acute{s}\else \'{s}\fi{}}}]{Adrich2005PRL}%
  \BibitemOpen
  \bibfield  {author} {\bibinfo {author} {\bibfnamefont {P.}~\bibnamefont
  {Adrich}}, \bibinfo {author} {\bibfnamefont {A.}~\bibnamefont {Klimkiewicz}},
  \bibinfo {author} {\bibfnamefont {M.}~\bibnamefont {Fallot}}, \bibinfo
  {author} {\bibfnamefont {K.}~\bibnamefont {Boretzky}}, \bibinfo {author}
  {\bibfnamefont {T.}~\bibnamefont {Aumann}}, \bibinfo {author} {\bibfnamefont
  {D.}~\bibnamefont {Cortina-Gil}}, \bibinfo {author} {\bibfnamefont {U.~D.}\
  \bibnamefont {Pramanik}}, \bibinfo {author} {\bibfnamefont {T.~W.}\
  \bibnamefont {Elze}}, \bibinfo {author} {\bibfnamefont {H.}~\bibnamefont
  {Emling}}, \bibinfo {author} {\bibfnamefont {H.}~\bibnamefont {Geissel}},
  \bibinfo {author} {\bibfnamefont {M.}~\bibnamefont {Hellstr\"om}}, \bibinfo
  {author} {\bibfnamefont {K.~L.}\ \bibnamefont {Jones}}, \bibinfo {author}
  {\bibfnamefont {J.~V.}\ \bibnamefont {Kratz}}, \bibinfo {author}
  {\bibfnamefont {R.}~\bibnamefont {Kulessa}}, \bibinfo {author} {\bibfnamefont
  {Y.}~\bibnamefont {Leifels}}, \bibinfo {author} {\bibfnamefont
  {C.}~\bibnamefont {Nociforo}}, \bibinfo {author} {\bibfnamefont
  {R.}~\bibnamefont {Palit}}, \bibinfo {author} {\bibfnamefont
  {H.}~\bibnamefont {Simon}}, \bibinfo {author} {\bibfnamefont
  {G.}~\bibnamefont {Sur\'owka}}, \bibinfo {author} {\bibfnamefont
  {K.}~\bibnamefont {S\"ummerer}}, \ and\ \bibinfo {author} {\bibfnamefont
  {W.}~\bibnamefont {Walu\ifmmode~\acute{s}\else \'{s}\fi{}}} (\bibinfo
  {collaboration} {LAND-FRS Collaboration}),\ }\href {\doibase
  10.1103/PhysRevLett.95.132501} {\bibfield  {journal} {\bibinfo  {journal}
  {Phys. Rev. Lett.}\ }\textbf {\bibinfo {volume} {95}},\ \bibinfo {pages}
  {132501} (\bibinfo {year} {2005})}\BibitemShut {NoStop}%
\bibitem [{\citenamefont {Stoitsov}\ \emph {et~al.}(2003)\citenamefont
  {Stoitsov}, \citenamefont {Dobaczewski}, \citenamefont {Nazarewicz},
  \citenamefont {Pittel},\ and\ \citenamefont {Dean}}]{Stoitsov2003PRC}%
  \BibitemOpen
  \bibfield  {author} {\bibinfo {author} {\bibfnamefont {M.~V.}\ \bibnamefont
  {Stoitsov}}, \bibinfo {author} {\bibfnamefont {J.}~\bibnamefont
  {Dobaczewski}}, \bibinfo {author} {\bibfnamefont {W.}~\bibnamefont
  {Nazarewicz}}, \bibinfo {author} {\bibfnamefont {S.}~\bibnamefont {Pittel}},
  \ and\ \bibinfo {author} {\bibfnamefont {D.~J.}\ \bibnamefont {Dean}},\
  }\href {\doibase 10.1103/PhysRevC.68.054312} {\bibfield  {journal} {\bibinfo
  {journal} {Phys. Rev. C}\ }\textbf {\bibinfo {volume} {68}},\ \bibinfo
  {pages} {054312} (\bibinfo {year} {2003})}\BibitemShut {NoStop}%
\bibitem [{\citenamefont {Erler}\ \emph {et~al.}(2012)\citenamefont {Erler},
  \citenamefont {Birge}, \citenamefont {Kortelainen}, \citenamefont
  {Nazarewicz}, \citenamefont {Olsen}, \citenamefont {Perhac},\ and\
  \citenamefont {Stoitsov}}]{Erler2012Nat}%
  \BibitemOpen
  \bibfield  {author} {\bibinfo {author} {\bibfnamefont {J.}~\bibnamefont
  {Erler}}, \bibinfo {author} {\bibfnamefont {N.}~\bibnamefont {Birge}},
  \bibinfo {author} {\bibfnamefont {M.}~\bibnamefont {Kortelainen}}, \bibinfo
  {author} {\bibfnamefont {W.}~\bibnamefont {Nazarewicz}}, \bibinfo {author}
  {\bibfnamefont {E.}~\bibnamefont {Olsen}}, \bibinfo {author} {\bibfnamefont
  {A.~M.}\ \bibnamefont {Perhac}}, \ and\ \bibinfo {author} {\bibfnamefont
  {M.}~\bibnamefont {Stoitsov}},\ }\href {\doibase 10.1038/nature11188}
  {\bibfield  {journal} {\bibinfo  {journal} {Nature}\ }\textbf {\bibinfo
  {volume} {486}},\ \bibinfo {pages} {509} (\bibinfo {year}
  {2012})}\BibitemShut {NoStop}%
\bibitem [{\citenamefont {Delaroche}\ \emph {et~al.}(2010)\citenamefont
  {Delaroche}, \citenamefont {Girod}, \citenamefont {Libert}, \citenamefont
  {Goutte}, \citenamefont {Hilaire}, \citenamefont {P\'eru}, \citenamefont
  {Pillet},\ and\ \citenamefont {Bertsch}}]{Delaroche2010PRC}%
  \BibitemOpen
  \bibfield  {author} {\bibinfo {author} {\bibfnamefont {J.~P.}\ \bibnamefont
  {Delaroche}}, \bibinfo {author} {\bibfnamefont {M.}~\bibnamefont {Girod}},
  \bibinfo {author} {\bibfnamefont {J.}~\bibnamefont {Libert}}, \bibinfo
  {author} {\bibfnamefont {H.}~\bibnamefont {Goutte}}, \bibinfo {author}
  {\bibfnamefont {S.}~\bibnamefont {Hilaire}}, \bibinfo {author} {\bibfnamefont
  {S.}~\bibnamefont {P\'eru}}, \bibinfo {author} {\bibfnamefont
  {N.}~\bibnamefont {Pillet}}, \ and\ \bibinfo {author} {\bibfnamefont {G.~F.}\
  \bibnamefont {Bertsch}},\ }\href {\doibase 10.1103/PhysRevC.81.014303}
  {\bibfield  {journal} {\bibinfo  {journal} {Phys. Rev. C}\ }\textbf {\bibinfo
  {volume} {81}},\ \bibinfo {pages} {014303} (\bibinfo {year}
  {2010})}\BibitemShut {NoStop}%
\bibitem [{\citenamefont {Afanasjev}\ \emph {et~al.}(2013)\citenamefont
  {Afanasjev}, \citenamefont {Agbemava}, \citenamefont {Ray},\ and\
  \citenamefont {Ring}}]{Afanasjev2013PLB}%
  \BibitemOpen
  \bibfield  {author} {\bibinfo {author} {\bibfnamefont {A.~V.}\ \bibnamefont
  {Afanasjev}}, \bibinfo {author} {\bibfnamefont {S.~E.}\ \bibnamefont
  {Agbemava}}, \bibinfo {author} {\bibfnamefont {D.}~\bibnamefont {Ray}}, \
  and\ \bibinfo {author} {\bibfnamefont {P.}~\bibnamefont {Ring}},\ }\href
  {\doibase https://doi.org/10.1016/j.physletb.2013.09.017} {\bibfield
  {journal} {\bibinfo  {journal} {Phys. Lett. B}\ }\textbf {\bibinfo {volume}
  {726}},\ \bibinfo {pages} {680 } (\bibinfo {year} {2013})}\BibitemShut
  {NoStop}%
\bibitem [{\citenamefont {Gridnev}\ \emph {et~al.}(2015)\citenamefont
  {Gridnev}, \citenamefont {Tarasov}, \citenamefont {Gridnev}, \citenamefont
  {Vi{\~{n}}as},\ and\ \citenamefont {Greiner}}]{Gridnev2015Book}%
  \BibitemOpen
  \bibfield  {author} {\bibinfo {author} {\bibfnamefont {K.~A.}\ \bibnamefont
  {Gridnev}}, \bibinfo {author} {\bibfnamefont {V.~N.}\ \bibnamefont
  {Tarasov}}, \bibinfo {author} {\bibfnamefont {D.~K.}\ \bibnamefont
  {Gridnev}}, \bibinfo {author} {\bibfnamefont {X.}~\bibnamefont
  {Vi{\~{n}}as}}, \ and\ \bibinfo {author} {\bibfnamefont {W.}~\bibnamefont
  {Greiner}},\ }\enquote {\bibinfo {title} {Stability {P}eninsulas in the
  {N}eutron-{R}ich {S}ector},}\ in\ \href {\doibase
  10.1007/978-3-319-10199-6_10} {\emph {\bibinfo {booktitle} {Nuclear Physics:
  Present and Future}}},\ \bibinfo {editor} {edited by\ \bibinfo {editor}
  {\bibfnamefont {W.}~\bibnamefont {Greiner}}}\ (\bibinfo  {publisher}
  {Springer International Publishing},\ \bibinfo {address} {Cham},\ \bibinfo
  {year} {2015})\ Chap.~\bibinfo {chapter} {10}, pp.\ \bibinfo {pages}
  {99--105}\BibitemShut {NoStop}%
\bibitem [{\citenamefont {Dobaczewski}\ \emph {et~al.}(1984)\citenamefont
  {Dobaczewski}, \citenamefont {Flocard},\ and\ \citenamefont
  {Treiner}}]{Dobaczewski1984NPA}%
  \BibitemOpen
  \bibfield  {author} {\bibinfo {author} {\bibfnamefont {J.}~\bibnamefont
  {Dobaczewski}}, \bibinfo {author} {\bibfnamefont {H.}~\bibnamefont
  {Flocard}}, \ and\ \bibinfo {author} {\bibfnamefont {J.}~\bibnamefont
  {Treiner}},\ }\href {\doibase https://doi.org/10.1016/0375-9474(84)90433-0}
  {\bibfield  {journal} {\bibinfo  {journal} {Nucl. Phys. A}\ }\textbf
  {\bibinfo {volume} {422}},\ \bibinfo {pages} {103} (\bibinfo {year}
  {1984})}\BibitemShut {NoStop}%
\bibitem [{\citenamefont {Nik\v{s}i\'{c}}\ \emph {et~al.}(2011)\citenamefont
  {Nik\v{s}i\'{c}}, \citenamefont {Vretenar},\ and\ \citenamefont
  {Ring}}]{Niksic2011PPNP}%
  \BibitemOpen
  \bibfield  {author} {\bibinfo {author} {\bibfnamefont {T.}~\bibnamefont
  {Nik\v{s}i\'{c}}}, \bibinfo {author} {\bibfnamefont {D.}~\bibnamefont
  {Vretenar}}, \ and\ \bibinfo {author} {\bibfnamefont {P.}~\bibnamefont
  {Ring}},\ }\href {\doibase 10.1016/j.ppnp.2011.01.055} {\bibfield  {journal}
  {\bibinfo  {journal} {Prog. Part. Nucl. Phys.}\ }\textbf {\bibinfo {volume}
  {66}},\ \bibinfo {pages} {519 } (\bibinfo {year} {2011})}\BibitemShut
  {NoStop}%
\bibitem [{\citenamefont {Dobaczewski}\ \emph {et~al.}(1996)\citenamefont
  {Dobaczewski}, \citenamefont {Nazarewicz}, \citenamefont {Werner},
  \citenamefont {Berger}, \citenamefont {Chinn},\ and\ \citenamefont
  {Decharg\'e}}]{Dobaczewski1996PRC}%
  \BibitemOpen
  \bibfield  {author} {\bibinfo {author} {\bibfnamefont {J.}~\bibnamefont
  {Dobaczewski}}, \bibinfo {author} {\bibfnamefont {W.}~\bibnamefont
  {Nazarewicz}}, \bibinfo {author} {\bibfnamefont {T.~R.}\ \bibnamefont
  {Werner}}, \bibinfo {author} {\bibfnamefont {J.~F.}\ \bibnamefont {Berger}},
  \bibinfo {author} {\bibfnamefont {C.~R.}\ \bibnamefont {Chinn}}, \ and\
  \bibinfo {author} {\bibfnamefont {J.}~\bibnamefont {Decharg\'e}},\ }\href
  {\doibase 10.1103/PhysRevC.53.2809} {\bibfield  {journal} {\bibinfo
  {journal} {Phys. Rev. C}\ }\textbf {\bibinfo {volume} {53}},\ \bibinfo
  {pages} {2809} (\bibinfo {year} {1996})}\BibitemShut {NoStop}%
\bibitem [{\citenamefont {Meng}(1998)}]{Meng1998NPA}%
  \BibitemOpen
  \bibfield  {author} {\bibinfo {author} {\bibfnamefont {J.}~\bibnamefont
  {Meng}},\ }\href {\doibase 10.1016/S0375-9474(98)00178-X} {\bibfield
  {journal} {\bibinfo  {journal} {Nucl. Phys. A}\ }\textbf {\bibinfo {volume}
  {635}},\ \bibinfo {pages} {3 } (\bibinfo {year} {1998})}\BibitemShut
  {NoStop}%
\bibitem [{\citenamefont {Zhou}\ \emph {et~al.}(2000)\citenamefont {Zhou},
  \citenamefont {Meng}, \citenamefont {Yamaji},\ and\ \citenamefont
  {Yang}}]{Zhou2000CPL}%
  \BibitemOpen
  \bibfield  {author} {\bibinfo {author} {\bibfnamefont {S.-G.}\ \bibnamefont
  {Zhou}}, \bibinfo {author} {\bibfnamefont {J.}~\bibnamefont {Meng}}, \bibinfo
  {author} {\bibfnamefont {S.}~\bibnamefont {Yamaji}}, \ and\ \bibinfo {author}
  {\bibfnamefont {S.-C.}\ \bibnamefont {Yang}},\ }\href {\doibase
  10.1088/0256-307x/17/10/006} {\bibfield  {journal} {\bibinfo  {journal}
  {Chin. Phys. Lett.}\ }\textbf {\bibinfo {volume} {17}},\ \bibinfo {pages}
  {717} (\bibinfo {year} {2000})}\BibitemShut {NoStop}%
\bibitem [{\citenamefont {Zhou}\ \emph {et~al.}(2003)\citenamefont {Zhou},
  \citenamefont {Meng},\ and\ \citenamefont {Ring}}]{Zhou2003PRC}%
  \BibitemOpen
  \bibfield  {author} {\bibinfo {author} {\bibfnamefont {S.-G.}\ \bibnamefont
  {Zhou}}, \bibinfo {author} {\bibfnamefont {J.}~\bibnamefont {Meng}}, \ and\
  \bibinfo {author} {\bibfnamefont {P.}~\bibnamefont {Ring}},\ }\href {\doibase
  10.1103/PhysRevC.68.034323} {\bibfield  {journal} {\bibinfo  {journal} {Phys.
  Rev. C}\ }\textbf {\bibinfo {volume} {68}},\ \bibinfo {pages} {034323}
  (\bibinfo {year} {2003})}\BibitemShut {NoStop}%
\bibitem [{\citenamefont {Stoitsov}\ \emph
  {et~al.}(1998{\natexlab{a}})\citenamefont {Stoitsov}, \citenamefont {Ring},
  \citenamefont {Vretenar},\ and\ \citenamefont
  {Lalazissis}}]{Stoitsov1998PRC_RHB}%
  \BibitemOpen
  \bibfield  {author} {\bibinfo {author} {\bibfnamefont {M.}~\bibnamefont
  {Stoitsov}}, \bibinfo {author} {\bibfnamefont {P.}~\bibnamefont {Ring}},
  \bibinfo {author} {\bibfnamefont {D.}~\bibnamefont {Vretenar}}, \ and\
  \bibinfo {author} {\bibfnamefont {G.~A.}\ \bibnamefont {Lalazissis}},\ }\href
  {\doibase 10.1103/PhysRevC.58.2086} {\bibfield  {journal} {\bibinfo
  {journal} {Phys. Rev. C}\ }\textbf {\bibinfo {volume} {58}},\ \bibinfo
  {pages} {2086} (\bibinfo {year} {1998}{\natexlab{a}})}\BibitemShut {NoStop}%
\bibitem [{\citenamefont {Stoitsov}\ \emph
  {et~al.}(1998{\natexlab{b}})\citenamefont {Stoitsov}, \citenamefont
  {Nazarewicz},\ and\ \citenamefont {Pittel}}]{Stoitsov1998PRC_HFB}%
  \BibitemOpen
  \bibfield  {author} {\bibinfo {author} {\bibfnamefont {M.~V.}\ \bibnamefont
  {Stoitsov}}, \bibinfo {author} {\bibfnamefont {W.}~\bibnamefont
  {Nazarewicz}}, \ and\ \bibinfo {author} {\bibfnamefont {S.}~\bibnamefont
  {Pittel}},\ }\href {\doibase 10.1103/PhysRevC.58.2092} {\bibfield  {journal}
  {\bibinfo  {journal} {Phys. Rev. C}\ }\textbf {\bibinfo {volume} {58}},\
  \bibinfo {pages} {2092} (\bibinfo {year} {1998}{\natexlab{b}})}\BibitemShut
  {NoStop}%
\bibitem [{\citenamefont {Zhang}\ \emph {et~al.}(2013)\citenamefont {Zhang},
  \citenamefont {Pei},\ and\ \citenamefont {Xu}}]{Zhang2013PRC}%
  \BibitemOpen
  \bibfield  {author} {\bibinfo {author} {\bibfnamefont {Y.~N.}\ \bibnamefont
  {Zhang}}, \bibinfo {author} {\bibfnamefont {J.~C.}\ \bibnamefont {Pei}}, \
  and\ \bibinfo {author} {\bibfnamefont {F.~R.}\ \bibnamefont {Xu}},\ }\href
  {\doibase 10.1103/PhysRevC.88.054305} {\bibfield  {journal} {\bibinfo
  {journal} {Phys. Rev. C}\ }\textbf {\bibinfo {volume} {88}},\ \bibinfo
  {pages} {054305} (\bibinfo {year} {2013})}\BibitemShut {NoStop}%
\bibitem [{\citenamefont {Meng}\ and\ \citenamefont
  {Ring}(1996)}]{Meng1996PRL}%
  \BibitemOpen
  \bibfield  {author} {\bibinfo {author} {\bibfnamefont {J.}~\bibnamefont
  {Meng}}\ and\ \bibinfo {author} {\bibfnamefont {P.}~\bibnamefont {Ring}},\
  }\href {\doibase 10.1103/PhysRevLett.77.3963} {\bibfield  {journal} {\bibinfo
   {journal} {Phys. Rev. Lett.}\ }\textbf {\bibinfo {volume} {77}},\ \bibinfo
  {pages} {3963} (\bibinfo {year} {1996})}\BibitemShut {NoStop}%
\bibitem [{\citenamefont {Meng}\ and\ \citenamefont
  {Ring}(1998)}]{Meng1998PRL}%
  \BibitemOpen
  \bibfield  {author} {\bibinfo {author} {\bibfnamefont {J.}~\bibnamefont
  {Meng}}\ and\ \bibinfo {author} {\bibfnamefont {P.}~\bibnamefont {Ring}},\
  }\href {\doibase 10.1103/PhysRevLett.80.460} {\bibfield  {journal} {\bibinfo
  {journal} {Phys. Rev. Lett.}\ }\textbf {\bibinfo {volume} {80}},\ \bibinfo
  {pages} {460} (\bibinfo {year} {1998})}\BibitemShut {NoStop}%
\bibitem [{\citenamefont {Meng}\ \emph {et~al.}(2002)\citenamefont {Meng},
  \citenamefont {Toki}, \citenamefont {Zeng}, \citenamefont {Zhang},\ and\
  \citenamefont {Zhou}}]{Meng2002PRC}%
  \BibitemOpen
  \bibfield  {author} {\bibinfo {author} {\bibfnamefont {J.}~\bibnamefont
  {Meng}}, \bibinfo {author} {\bibfnamefont {H.}~\bibnamefont {Toki}}, \bibinfo
  {author} {\bibfnamefont {J.~Y.}\ \bibnamefont {Zeng}}, \bibinfo {author}
  {\bibfnamefont {S.~Q.}\ \bibnamefont {Zhang}}, \ and\ \bibinfo {author}
  {\bibfnamefont {S.-G.}\ \bibnamefont {Zhou}},\ }\href {\doibase
  10.1103/PhysRevC.65.041302} {\bibfield  {journal} {\bibinfo  {journal} {Phys.
  Rev. C}\ }\textbf {\bibinfo {volume} {65}},\ \bibinfo {pages} {041302}
  (\bibinfo {year} {2002})}\BibitemShut {NoStop}%
\bibitem [{\citenamefont {Zhang}\ \emph {et~al.}(2002)\citenamefont {Zhang},
  \citenamefont {Meng}, \citenamefont {Zhou},\ and\ \citenamefont
  {Zeng}}]{Zhang2002CPL}%
  \BibitemOpen
  \bibfield  {author} {\bibinfo {author} {\bibfnamefont {S.-Q.}\ \bibnamefont
  {Zhang}}, \bibinfo {author} {\bibfnamefont {J.}~\bibnamefont {Meng}},
  \bibinfo {author} {\bibfnamefont {S.-G.}\ \bibnamefont {Zhou}}, \ and\
  \bibinfo {author} {\bibfnamefont {J.-Y.}\ \bibnamefont {Zeng}},\ }\href
  {\doibase 10.1088/0256-307x/19/3/308} {\bibfield  {journal} {\bibinfo
  {journal} {Chin. Phys. Lett.}\ }\textbf {\bibinfo {volume} {19}},\ \bibinfo
  {pages} {312} (\bibinfo {year} {2002})}\BibitemShut {NoStop}%
\bibitem [{\citenamefont {Meng}\ \emph {et~al.}(1998)\citenamefont {Meng},
  \citenamefont {Sugawara-Tanabe}, \citenamefont {Yamaji}, \citenamefont
  {Ring},\ and\ \citenamefont {Arima}}]{Meng1998PRC}%
  \BibitemOpen
  \bibfield  {author} {\bibinfo {author} {\bibfnamefont {J.}~\bibnamefont
  {Meng}}, \bibinfo {author} {\bibfnamefont {K.}~\bibnamefont
  {Sugawara-Tanabe}}, \bibinfo {author} {\bibfnamefont {S.}~\bibnamefont
  {Yamaji}}, \bibinfo {author} {\bibfnamefont {P.}~\bibnamefont {Ring}}, \ and\
  \bibinfo {author} {\bibfnamefont {A.}~\bibnamefont {Arima}},\ }\href
  {\doibase 10.1103/PhysRevC.58.R628} {\bibfield  {journal} {\bibinfo
  {journal} {Phys. Rev. C}\ }\textbf {\bibinfo {volume} {58}},\ \bibinfo
  {pages} {R628} (\bibinfo {year} {1998})}\BibitemShut {NoStop}%
\bibitem [{\citenamefont {Meng}\ \emph {et~al.}(1999)\citenamefont {Meng},
  \citenamefont {Sugawara-Tanabe}, \citenamefont {Yamaji},\ and\ \citenamefont
  {Arima}}]{Meng1999PRC}%
  \BibitemOpen
  \bibfield  {author} {\bibinfo {author} {\bibfnamefont {J.}~\bibnamefont
  {Meng}}, \bibinfo {author} {\bibfnamefont {K.}~\bibnamefont
  {Sugawara-Tanabe}}, \bibinfo {author} {\bibfnamefont {S.}~\bibnamefont
  {Yamaji}}, \ and\ \bibinfo {author} {\bibfnamefont {A.}~\bibnamefont
  {Arima}},\ }\href {\doibase 10.1103/PhysRevC.59.154} {\bibfield  {journal}
  {\bibinfo  {journal} {Phys. Rev. C}\ }\textbf {\bibinfo {volume} {59}},\
  \bibinfo {pages} {154} (\bibinfo {year} {1999})}\BibitemShut {NoStop}%
\bibitem [{\citenamefont {Zhang}\ \emph {et~al.}(2005)\citenamefont {Zhang},
  \citenamefont {Meng}, \citenamefont {Zhang}, \citenamefont {Geng},\ and\
  \citenamefont {Toki}}]{Zhang2005NPA}%
  \BibitemOpen
  \bibfield  {author} {\bibinfo {author} {\bibfnamefont {W.}~\bibnamefont
  {Zhang}}, \bibinfo {author} {\bibfnamefont {J.}~\bibnamefont {Meng}},
  \bibinfo {author} {\bibfnamefont {S.~Q.}\ \bibnamefont {Zhang}}, \bibinfo
  {author} {\bibfnamefont {L.~S.}\ \bibnamefont {Geng}}, \ and\ \bibinfo
  {author} {\bibfnamefont {H.}~\bibnamefont {Toki}},\ }\href {\doibase
  10.1016/j.nuclphysa.2005.02.086} {\bibfield  {journal} {\bibinfo  {journal}
  {Nucl. Phys. A}\ }\textbf {\bibinfo {volume} {753}},\ \bibinfo {pages} {106 }
  (\bibinfo {year} {2005})}\BibitemShut {NoStop}%
\bibitem [{\citenamefont {Xia}\ \emph {et~al.}(2018)\citenamefont {Xia},
  \citenamefont {Lim}, \citenamefont {Zhao}, \citenamefont {Liang},
  \citenamefont {Qu}, \citenamefont {Chen}, \citenamefont {Liu}, \citenamefont
  {Zhang}, \citenamefont {Zhang}, \citenamefont {Kim},\ and\ \citenamefont
  {Meng}}]{Xia2018ADNDT}%
  \BibitemOpen
  \bibfield  {author} {\bibinfo {author} {\bibfnamefont {X.~W.}\ \bibnamefont
  {Xia}}, \bibinfo {author} {\bibfnamefont {Y.}~\bibnamefont {Lim}}, \bibinfo
  {author} {\bibfnamefont {P.~W.}\ \bibnamefont {Zhao}}, \bibinfo {author}
  {\bibfnamefont {H.~Z.}\ \bibnamefont {Liang}}, \bibinfo {author}
  {\bibfnamefont {X.~Y.}\ \bibnamefont {Qu}}, \bibinfo {author} {\bibfnamefont
  {Y.}~\bibnamefont {Chen}}, \bibinfo {author} {\bibfnamefont {H.}~\bibnamefont
  {Liu}}, \bibinfo {author} {\bibfnamefont {L.~F.}\ \bibnamefont {Zhang}},
  \bibinfo {author} {\bibfnamefont {S.~Q.}\ \bibnamefont {Zhang}}, \bibinfo
  {author} {\bibfnamefont {Y.}~\bibnamefont {Kim}}, \ and\ \bibinfo {author}
  {\bibfnamefont {J.}~\bibnamefont {Meng}},\ }\href {\doibase
  10.1016/j.adt.2017.09.001} {\bibfield  {journal} {\bibinfo  {journal} {At.
  Data Nucl. Data Tables}\ }\textbf {\bibinfo {volume} {121-122}},\ \bibinfo
  {pages} {1 } (\bibinfo {year} {2018})}\BibitemShut {NoStop}%
\bibitem [{\citenamefont {Zhou}\ \emph {et~al.}(2010)\citenamefont {Zhou},
  \citenamefont {Meng}, \citenamefont {Ring},\ and\ \citenamefont
  {Zhao}}]{Zhou2010PRC}%
  \BibitemOpen
  \bibfield  {author} {\bibinfo {author} {\bibfnamefont {S.-G.}\ \bibnamefont
  {Zhou}}, \bibinfo {author} {\bibfnamefont {J.}~\bibnamefont {Meng}}, \bibinfo
  {author} {\bibfnamefont {P.}~\bibnamefont {Ring}}, \ and\ \bibinfo {author}
  {\bibfnamefont {E.-G.}\ \bibnamefont {Zhao}},\ }\href {\doibase
  10.1103/PhysRevC.82.011301} {\bibfield  {journal} {\bibinfo  {journal} {Phys.
  Rev. C}\ }\textbf {\bibinfo {volume} {82}},\ \bibinfo {pages} {011301}
  (\bibinfo {year} {2010})}\BibitemShut {NoStop}%
\bibitem [{\citenamefont {Li}\ \emph {et~al.}(2012)\citenamefont {Li},
  \citenamefont {Meng}, \citenamefont {Ring}, \citenamefont {Zhao},\ and\
  \citenamefont {Zhou}}]{Li2012PRC}%
  \BibitemOpen
  \bibfield  {author} {\bibinfo {author} {\bibfnamefont {L.}~\bibnamefont
  {Li}}, \bibinfo {author} {\bibfnamefont {J.}~\bibnamefont {Meng}}, \bibinfo
  {author} {\bibfnamefont {P.}~\bibnamefont {Ring}}, \bibinfo {author}
  {\bibfnamefont {E.-G.}\ \bibnamefont {Zhao}}, \ and\ \bibinfo {author}
  {\bibfnamefont {S.-G.}\ \bibnamefont {Zhou}},\ }\href {\doibase
  10.1103/PhysRevC.85.024312} {\bibfield  {journal} {\bibinfo  {journal} {Phys.
  Rev. C}\ }\textbf {\bibinfo {volume} {85}},\ \bibinfo {pages} {024312}
  (\bibinfo {year} {2012})}\BibitemShut {NoStop}%
\bibitem [{\citenamefont {Sun}\ and\ \citenamefont {Zhou}(2021)}]{Sun2021SciB}%
  \BibitemOpen
  \bibfield  {author} {\bibinfo {author} {\bibfnamefont {X.-X.}\ \bibnamefont
  {Sun}}\ and\ \bibinfo {author} {\bibfnamefont {S.-G.}\ \bibnamefont {Zhou}},\
  }\href {https://doi.org/10.1016/j.scib.2021.07.005} {\bibfield  {journal}
  {\bibinfo  {journal} {Sci. Bull.}\ ,\ \bibinfo {pages}
  {doi:10.1016/j.scib.2021.07.005}} (\bibinfo {year} {2021})}\BibitemShut
  {NoStop}%
\bibitem [{\citenamefont {Sun}\ \emph {et~al.}(2018)\citenamefont {Sun},
  \citenamefont {Zhao},\ and\ \citenamefont {Zhou}}]{Sun2018PLB}%
  \BibitemOpen
  \bibfield  {author} {\bibinfo {author} {\bibfnamefont {X.-X.}\ \bibnamefont
  {Sun}}, \bibinfo {author} {\bibfnamefont {J.}~\bibnamefont {Zhao}}, \ and\
  \bibinfo {author} {\bibfnamefont {S.-G.}\ \bibnamefont {Zhou}},\ }\href
  {\doibase 10.1016/j.physletb.2018.08.071} {\bibfield  {journal} {\bibinfo
  {journal} {Phys. Lett. B}\ }\textbf {\bibinfo {volume} {785}},\ \bibinfo
  {pages} {530 } (\bibinfo {year} {2018})}\BibitemShut {NoStop}%
\bibitem [{\citenamefont {Sun}\ \emph {et~al.}(2020)\citenamefont {Sun},
  \citenamefont {Zhao},\ and\ \citenamefont {Zhou}}]{Sun2020NPA}%
  \BibitemOpen
  \bibfield  {author} {\bibinfo {author} {\bibfnamefont {X.-X.}\ \bibnamefont
  {Sun}}, \bibinfo {author} {\bibfnamefont {J.}~\bibnamefont {Zhao}}, \ and\
  \bibinfo {author} {\bibfnamefont {S.-G.}\ \bibnamefont {Zhou}},\ }\href
  {\doibase https://doi.org/10.1016/j.nuclphysa.2020.122011} {\bibfield
  {journal} {\bibinfo  {journal} {Nucl. Phys. A}\ }\textbf {\bibinfo {volume}
  {1003}},\ \bibinfo {pages} {122011} (\bibinfo {year} {2020})}\BibitemShut
  {NoStop}%
\bibitem [{\citenamefont {Zhang}\ \emph {et~al.}(2019)\citenamefont {Zhang},
  \citenamefont {Wang},\ and\ \citenamefont {Zhang}}]{Zhang2019PRC}%
  \BibitemOpen
  \bibfield  {author} {\bibinfo {author} {\bibfnamefont {K.~Y.}\ \bibnamefont
  {Zhang}}, \bibinfo {author} {\bibfnamefont {D.~Y.}\ \bibnamefont {Wang}}, \
  and\ \bibinfo {author} {\bibfnamefont {S.~Q.}\ \bibnamefont {Zhang}},\ }\href
  {\doibase 10.1103/PhysRevC.100.034312} {\bibfield  {journal} {\bibinfo
  {journal} {Phys. Rev. C}\ }\textbf {\bibinfo {volume} {100}},\ \bibinfo
  {pages} {034312} (\bibinfo {year} {2019})}\BibitemShut {NoStop}%
\bibitem [{\citenamefont {Yang}\ \emph
  {et~al.}(2021{\natexlab{a}})\citenamefont {Yang}, \citenamefont {Kubota},
  \citenamefont {Corsi}, \citenamefont {Yoshida}, \citenamefont {Sun},
  \citenamefont {Li}, \citenamefont {Kimura}, \citenamefont {Michel},
  \citenamefont {Ogata}, \citenamefont {Yuan}, \citenamefont {Yuan},
  \citenamefont {Authelet}, \citenamefont {Baba}, \citenamefont {Caesar},
  \citenamefont {Calvet}, \citenamefont {Delbart}, \citenamefont {Dozono},
  \citenamefont {Feng}, \citenamefont {Flavigny}, \citenamefont {Gheller},
  \citenamefont {Gibelin}, \citenamefont {Giganon}, \citenamefont {Gillibert},
  \citenamefont {Hasegawa}, \citenamefont {Isobe}, \citenamefont {Kanaya},
  \citenamefont {Kawakami}, \citenamefont {Kim}, \citenamefont {Kiyokawa},
  \citenamefont {Kobayashi}, \citenamefont {Kobayashi}, \citenamefont
  {Kobayashi}, \citenamefont {Kondo}, \citenamefont {Korkulu}, \citenamefont
  {Koyama}, \citenamefont {Lapoux}, \citenamefont {Maeda}, \citenamefont
  {Marqu\'es}, \citenamefont {Motobayashi}, \citenamefont {Miyazaki},
  \citenamefont {Nakamura}, \citenamefont {Nakatsuka}, \citenamefont {Nishio},
  \citenamefont {Obertelli}, \citenamefont {Ohkura}, \citenamefont {Orr},
  \citenamefont {Ota}, \citenamefont {Otsu}, \citenamefont {Ozaki},
  \citenamefont {Panin}, \citenamefont {Paschalis}, \citenamefont {Pollacco},
  \citenamefont {Reichert}, \citenamefont {Rouss\'e}, \citenamefont {Saito},
  \citenamefont {Sakaguchi}, \citenamefont {Sako}, \citenamefont {Santamaria},
  \citenamefont {Sasano}, \citenamefont {Sato}, \citenamefont {Shikata},
  \citenamefont {Shimizu}, \citenamefont {Shindo}, \citenamefont {Stuhl},
  \citenamefont {Sumikama}, \citenamefont {Sun}, \citenamefont {Tabata},
  \citenamefont {Togano}, \citenamefont {Tsubota}, \citenamefont {Xu},
  \citenamefont {Yasuda}, \citenamefont {Yoneda}, \citenamefont {Zenihiro},
  \citenamefont {Zhou}, \citenamefont {Zuo},\ and\ \citenamefont
  {Uesaka}}]{Yang2021PRL}%
  \BibitemOpen
  \bibfield  {author} {\bibinfo {author} {\bibfnamefont {Z.~H.}\ \bibnamefont
  {Yang}}, \bibinfo {author} {\bibfnamefont {Y.}~\bibnamefont {Kubota}},
  \bibinfo {author} {\bibfnamefont {A.}~\bibnamefont {Corsi}}, \bibinfo
  {author} {\bibfnamefont {K.}~\bibnamefont {Yoshida}}, \bibinfo {author}
  {\bibfnamefont {X.-X.}\ \bibnamefont {Sun}}, \bibinfo {author} {\bibfnamefont
  {J.~G.}\ \bibnamefont {Li}}, \bibinfo {author} {\bibfnamefont
  {M.}~\bibnamefont {Kimura}}, \bibinfo {author} {\bibfnamefont
  {N.}~\bibnamefont {Michel}}, \bibinfo {author} {\bibfnamefont
  {K.}~\bibnamefont {Ogata}}, \bibinfo {author} {\bibfnamefont {C.~X.}\
  \bibnamefont {Yuan}}, \bibinfo {author} {\bibfnamefont {Q.}~\bibnamefont
  {Yuan}}, \bibinfo {author} {\bibfnamefont {G.}~\bibnamefont {Authelet}},
  \bibinfo {author} {\bibfnamefont {H.}~\bibnamefont {Baba}}, \bibinfo {author}
  {\bibfnamefont {C.}~\bibnamefont {Caesar}}, \bibinfo {author} {\bibfnamefont
  {D.}~\bibnamefont {Calvet}}, \bibinfo {author} {\bibfnamefont
  {A.}~\bibnamefont {Delbart}}, \bibinfo {author} {\bibfnamefont
  {M.}~\bibnamefont {Dozono}}, \bibinfo {author} {\bibfnamefont
  {J.}~\bibnamefont {Feng}}, \bibinfo {author} {\bibfnamefont {F.}~\bibnamefont
  {Flavigny}}, \bibinfo {author} {\bibfnamefont {J.-M.}\ \bibnamefont
  {Gheller}}, \bibinfo {author} {\bibfnamefont {J.}~\bibnamefont {Gibelin}},
  \bibinfo {author} {\bibfnamefont {A.}~\bibnamefont {Giganon}}, \bibinfo
  {author} {\bibfnamefont {A.}~\bibnamefont {Gillibert}}, \bibinfo {author}
  {\bibfnamefont {K.}~\bibnamefont {Hasegawa}}, \bibinfo {author}
  {\bibfnamefont {T.}~\bibnamefont {Isobe}}, \bibinfo {author} {\bibfnamefont
  {Y.}~\bibnamefont {Kanaya}}, \bibinfo {author} {\bibfnamefont
  {S.}~\bibnamefont {Kawakami}}, \bibinfo {author} {\bibfnamefont
  {D.}~\bibnamefont {Kim}}, \bibinfo {author} {\bibfnamefont {Y.}~\bibnamefont
  {Kiyokawa}}, \bibinfo {author} {\bibfnamefont {M.}~\bibnamefont {Kobayashi}},
  \bibinfo {author} {\bibfnamefont {N.}~\bibnamefont {Kobayashi}}, \bibinfo
  {author} {\bibfnamefont {T.}~\bibnamefont {Kobayashi}}, \bibinfo {author}
  {\bibfnamefont {Y.}~\bibnamefont {Kondo}}, \bibinfo {author} {\bibfnamefont
  {Z.}~\bibnamefont {Korkulu}}, \bibinfo {author} {\bibfnamefont
  {S.}~\bibnamefont {Koyama}}, \bibinfo {author} {\bibfnamefont
  {V.}~\bibnamefont {Lapoux}}, \bibinfo {author} {\bibfnamefont
  {Y.}~\bibnamefont {Maeda}}, \bibinfo {author} {\bibfnamefont {F.~M.}\
  \bibnamefont {Marqu\'es}}, \bibinfo {author} {\bibfnamefont {T.}~\bibnamefont
  {Motobayashi}}, \bibinfo {author} {\bibfnamefont {T.}~\bibnamefont
  {Miyazaki}}, \bibinfo {author} {\bibfnamefont {T.}~\bibnamefont {Nakamura}},
  \bibinfo {author} {\bibfnamefont {N.}~\bibnamefont {Nakatsuka}}, \bibinfo
  {author} {\bibfnamefont {Y.}~\bibnamefont {Nishio}}, \bibinfo {author}
  {\bibfnamefont {A.}~\bibnamefont {Obertelli}}, \bibinfo {author}
  {\bibfnamefont {A.}~\bibnamefont {Ohkura}}, \bibinfo {author} {\bibfnamefont
  {N.~A.}\ \bibnamefont {Orr}}, \bibinfo {author} {\bibfnamefont
  {S.}~\bibnamefont {Ota}}, \bibinfo {author} {\bibfnamefont {H.}~\bibnamefont
  {Otsu}}, \bibinfo {author} {\bibfnamefont {T.}~\bibnamefont {Ozaki}},
  \bibinfo {author} {\bibfnamefont {V.}~\bibnamefont {Panin}}, \bibinfo
  {author} {\bibfnamefont {S.}~\bibnamefont {Paschalis}}, \bibinfo {author}
  {\bibfnamefont {E.~C.}\ \bibnamefont {Pollacco}}, \bibinfo {author}
  {\bibfnamefont {S.}~\bibnamefont {Reichert}}, \bibinfo {author}
  {\bibfnamefont {J.-Y.}\ \bibnamefont {Rouss\'e}}, \bibinfo {author}
  {\bibfnamefont {A.~T.}\ \bibnamefont {Saito}}, \bibinfo {author}
  {\bibfnamefont {S.}~\bibnamefont {Sakaguchi}}, \bibinfo {author}
  {\bibfnamefont {M.}~\bibnamefont {Sako}}, \bibinfo {author} {\bibfnamefont
  {C.}~\bibnamefont {Santamaria}}, \bibinfo {author} {\bibfnamefont
  {M.}~\bibnamefont {Sasano}}, \bibinfo {author} {\bibfnamefont
  {H.}~\bibnamefont {Sato}}, \bibinfo {author} {\bibfnamefont {M.}~\bibnamefont
  {Shikata}}, \bibinfo {author} {\bibfnamefont {Y.}~\bibnamefont {Shimizu}},
  \bibinfo {author} {\bibfnamefont {Y.}~\bibnamefont {Shindo}}, \bibinfo
  {author} {\bibfnamefont {L.}~\bibnamefont {Stuhl}}, \bibinfo {author}
  {\bibfnamefont {T.}~\bibnamefont {Sumikama}}, \bibinfo {author}
  {\bibfnamefont {Y.~L.}\ \bibnamefont {Sun}}, \bibinfo {author} {\bibfnamefont
  {M.}~\bibnamefont {Tabata}}, \bibinfo {author} {\bibfnamefont
  {Y.}~\bibnamefont {Togano}}, \bibinfo {author} {\bibfnamefont
  {J.}~\bibnamefont {Tsubota}}, \bibinfo {author} {\bibfnamefont {F.~R.}\
  \bibnamefont {Xu}}, \bibinfo {author} {\bibfnamefont {J.}~\bibnamefont
  {Yasuda}}, \bibinfo {author} {\bibfnamefont {K.}~\bibnamefont {Yoneda}},
  \bibinfo {author} {\bibfnamefont {J.}~\bibnamefont {Zenihiro}}, \bibinfo
  {author} {\bibfnamefont {S.-G.}\ \bibnamefont {Zhou}}, \bibinfo {author}
  {\bibfnamefont {W.}~\bibnamefont {Zuo}}, \ and\ \bibinfo {author}
  {\bibfnamefont {T.}~\bibnamefont {Uesaka}},\ }\href {\doibase
  10.1103/PhysRevLett.126.082501} {\bibfield  {journal} {\bibinfo  {journal}
  {Phys. Rev. Lett.}\ }\textbf {\bibinfo {volume} {126}},\ \bibinfo {pages}
  {082501} (\bibinfo {year} {2021}{\natexlab{a}})}\BibitemShut {NoStop}%
\bibitem [{\citenamefont {Wang}\ \emph {et~al.}(2021)\citenamefont {Wang},
  \citenamefont {Huang}, \citenamefont {Kondev}, \citenamefont {Audi},\ and\
  \citenamefont {Naimi}}]{Wang2021CPC}%
  \BibitemOpen
  \bibfield  {author} {\bibinfo {author} {\bibfnamefont {M.}~\bibnamefont
  {Wang}}, \bibinfo {author} {\bibfnamefont {W.~J.}\ \bibnamefont {Huang}},
  \bibinfo {author} {\bibfnamefont {F.~G.}\ \bibnamefont {Kondev}}, \bibinfo
  {author} {\bibfnamefont {G.}~\bibnamefont {Audi}}, \ and\ \bibinfo {author}
  {\bibfnamefont {S.}~\bibnamefont {Naimi}},\ }\href {\doibase
  10.1088/1674-1137/abddaf} {\bibfield  {journal} {\bibinfo  {journal} {Chin.
  Phys. C}\ }\textbf {\bibinfo {volume} {45}},\ \bibinfo {pages} {030003}
  (\bibinfo {year} {2021})}\BibitemShut {NoStop}%
\bibitem [{\citenamefont {Zhao}\ \emph {et~al.}(2010)\citenamefont {Zhao},
  \citenamefont {Li}, \citenamefont {Yao},\ and\ \citenamefont
  {Meng}}]{Zhao2010PRC}%
  \BibitemOpen
  \bibfield  {author} {\bibinfo {author} {\bibfnamefont {P.~W.}\ \bibnamefont
  {Zhao}}, \bibinfo {author} {\bibfnamefont {Z.~P.}\ \bibnamefont {Li}},
  \bibinfo {author} {\bibfnamefont {J.~M.}\ \bibnamefont {Yao}}, \ and\
  \bibinfo {author} {\bibfnamefont {J.}~\bibnamefont {Meng}},\ }\href {\doibase
  10.1103/PhysRevC.82.054319} {\bibfield  {journal} {\bibinfo  {journal} {Phys.
  Rev. C}\ }\textbf {\bibinfo {volume} {82}},\ \bibinfo {pages} {054319}
  (\bibinfo {year} {2010})}\BibitemShut {NoStop}%
\bibitem [{\citenamefont {Zhang}\ \emph {et~al.}(2021)\citenamefont {Zhang},
  \citenamefont {He}, \citenamefont {Meng}, \citenamefont {Pan}, \citenamefont
  {Shen}, \citenamefont {Wang},\ and\ \citenamefont {Zhang}}]{Zhang2021PRC}%
  \BibitemOpen
  \bibfield  {author} {\bibinfo {author} {\bibfnamefont {K.}~\bibnamefont
  {Zhang}}, \bibinfo {author} {\bibfnamefont {X.}~\bibnamefont {He}}, \bibinfo
  {author} {\bibfnamefont {J.}~\bibnamefont {Meng}}, \bibinfo {author}
  {\bibfnamefont {C.}~\bibnamefont {Pan}}, \bibinfo {author} {\bibfnamefont
  {C.}~\bibnamefont {Shen}}, \bibinfo {author} {\bibfnamefont {C.}~\bibnamefont
  {Wang}}, \ and\ \bibinfo {author} {\bibfnamefont {S.}~\bibnamefont {Zhang}},\
  }\href {\doibase 10.1103/PhysRevC.104.L021301} {\bibfield  {journal}
  {\bibinfo  {journal} {Phys. Rev. C}\ }\textbf {\bibinfo {volume} {104}},\
  \bibinfo {pages} {L021301} (\bibinfo {year} {2021})}\BibitemShut {NoStop}%
\bibitem [{\citenamefont {Zhang}\ \emph {et~al.}(2020)\citenamefont {Zhang},
  \citenamefont {Cheoun}, \citenamefont {Choi}, \citenamefont {Chong},
  \citenamefont {Dong}, \citenamefont {Geng}, \citenamefont {Ha}, \citenamefont
  {He}, \citenamefont {Heo}, \citenamefont {Ho}, \citenamefont {In},
  \citenamefont {Kim}, \citenamefont {Kim}, \citenamefont {Lee}, \citenamefont
  {Lee}, \citenamefont {Li}, \citenamefont {Luo}, \citenamefont {Meng},
  \citenamefont {Mun}, \citenamefont {Niu}, \citenamefont {Pan}, \citenamefont
  {Papakonstantinou}, \citenamefont {Shang}, \citenamefont {Shen},
  \citenamefont {Shen}, \citenamefont {Sun}, \citenamefont {Sun}, \citenamefont
  {Tam}, \citenamefont {Thaivayongnou}, \citenamefont {Wang}, \citenamefont
  {Wong}, \citenamefont {Xia}, \citenamefont {Yan}, \citenamefont {Yeung},
  \citenamefont {Yiu}, \citenamefont {Zhang}, \citenamefont {Zhang},\ and\
  \citenamefont {Zhou}}]{Zhang2020PRC}%
  \BibitemOpen
  \bibfield  {author} {\bibinfo {author} {\bibfnamefont {K.}~\bibnamefont
  {Zhang}}, \bibinfo {author} {\bibfnamefont {M.-K.}\ \bibnamefont {Cheoun}},
  \bibinfo {author} {\bibfnamefont {Y.-B.}\ \bibnamefont {Choi}}, \bibinfo
  {author} {\bibfnamefont {P.~S.}\ \bibnamefont {Chong}}, \bibinfo {author}
  {\bibfnamefont {J.}~\bibnamefont {Dong}}, \bibinfo {author} {\bibfnamefont
  {L.}~\bibnamefont {Geng}}, \bibinfo {author} {\bibfnamefont {E.}~\bibnamefont
  {Ha}}, \bibinfo {author} {\bibfnamefont {X.}~\bibnamefont {He}}, \bibinfo
  {author} {\bibfnamefont {C.}~\bibnamefont {Heo}}, \bibinfo {author}
  {\bibfnamefont {M.~C.}\ \bibnamefont {Ho}}, \bibinfo {author} {\bibfnamefont
  {E.~J.}\ \bibnamefont {In}}, \bibinfo {author} {\bibfnamefont
  {S.}~\bibnamefont {Kim}}, \bibinfo {author} {\bibfnamefont {Y.}~\bibnamefont
  {Kim}}, \bibinfo {author} {\bibfnamefont {C.-H.}\ \bibnamefont {Lee}},
  \bibinfo {author} {\bibfnamefont {J.}~\bibnamefont {Lee}}, \bibinfo {author}
  {\bibfnamefont {Z.}~\bibnamefont {Li}}, \bibinfo {author} {\bibfnamefont
  {T.}~\bibnamefont {Luo}}, \bibinfo {author} {\bibfnamefont {J.}~\bibnamefont
  {Meng}}, \bibinfo {author} {\bibfnamefont {M.-H.}\ \bibnamefont {Mun}},
  \bibinfo {author} {\bibfnamefont {Z.}~\bibnamefont {Niu}}, \bibinfo {author}
  {\bibfnamefont {C.}~\bibnamefont {Pan}}, \bibinfo {author} {\bibfnamefont
  {P.}~\bibnamefont {Papakonstantinou}}, \bibinfo {author} {\bibfnamefont
  {X.}~\bibnamefont {Shang}}, \bibinfo {author} {\bibfnamefont
  {C.}~\bibnamefont {Shen}}, \bibinfo {author} {\bibfnamefont {G.}~\bibnamefont
  {Shen}}, \bibinfo {author} {\bibfnamefont {W.}~\bibnamefont {Sun}}, \bibinfo
  {author} {\bibfnamefont {X.-X.}\ \bibnamefont {Sun}}, \bibinfo {author}
  {\bibfnamefont {C.~K.}\ \bibnamefont {Tam}}, \bibinfo {author} {\bibnamefont
  {Thaivayongnou}}, \bibinfo {author} {\bibfnamefont {C.}~\bibnamefont {Wang}},
  \bibinfo {author} {\bibfnamefont {S.~H.}\ \bibnamefont {Wong}}, \bibinfo
  {author} {\bibfnamefont {X.}~\bibnamefont {Xia}}, \bibinfo {author}
  {\bibfnamefont {Y.}~\bibnamefont {Yan}}, \bibinfo {author} {\bibfnamefont
  {R.~W.-Y.}\ \bibnamefont {Yeung}}, \bibinfo {author} {\bibfnamefont {T.~C.}\
  \bibnamefont {Yiu}}, \bibinfo {author} {\bibfnamefont {S.}~\bibnamefont
  {Zhang}}, \bibinfo {author} {\bibfnamefont {W.}~\bibnamefont {Zhang}}, \ and\
  \bibinfo {author} {\bibfnamefont {S.-G.}\ \bibnamefont {Zhou}} (\bibinfo
  {collaboration} {DRHBc Mass Table Collaboration}),\ }\href
  {https://link.aps.org/doi/10.1103/PhysRevC.102.024314} {\bibfield  {journal}
  {\bibinfo  {journal} {Phys. Rev. C}\ }\textbf {\bibinfo {volume} {102}},\
  \bibinfo {pages} {024314} (\bibinfo {year} {2020})}\BibitemShut {NoStop}%
\bibitem [{\citenamefont {Zhao}\ \emph {et~al.}(2012)\citenamefont {Zhao},
  \citenamefont {Song}, \citenamefont {Sun}, \citenamefont {Geissel},\ and\
  \citenamefont {Meng}}]{Zhao2012PRCmass}%
  \BibitemOpen
  \bibfield  {author} {\bibinfo {author} {\bibfnamefont {P.~W.}\ \bibnamefont
  {Zhao}}, \bibinfo {author} {\bibfnamefont {L.~S.}\ \bibnamefont {Song}},
  \bibinfo {author} {\bibfnamefont {B.}~\bibnamefont {Sun}}, \bibinfo {author}
  {\bibfnamefont {H.}~\bibnamefont {Geissel}}, \ and\ \bibinfo {author}
  {\bibfnamefont {J.}~\bibnamefont {Meng}},\ }\href {\doibase
  10.1103/PhysRevC.86.064324} {\bibfield  {journal} {\bibinfo  {journal} {Phys.
  Rev. C}\ }\textbf {\bibinfo {volume} {86}},\ \bibinfo {pages} {064324}
  (\bibinfo {year} {2012})}\BibitemShut {NoStop}%
\bibitem [{\citenamefont {Lu}\ \emph {et~al.}(2015)\citenamefont {Lu},
  \citenamefont {Li}, \citenamefont {Li}, \citenamefont {Yao},\ and\
  \citenamefont {Meng}}]{Lu2015PRC}%
  \BibitemOpen
  \bibfield  {author} {\bibinfo {author} {\bibfnamefont {K.~Q.}\ \bibnamefont
  {Lu}}, \bibinfo {author} {\bibfnamefont {Z.~X.}\ \bibnamefont {Li}}, \bibinfo
  {author} {\bibfnamefont {Z.~P.}\ \bibnamefont {Li}}, \bibinfo {author}
  {\bibfnamefont {J.~M.}\ \bibnamefont {Yao}}, \ and\ \bibinfo {author}
  {\bibfnamefont {J.}~\bibnamefont {Meng}},\ }\href {\doibase
  10.1103/PhysRevC.91.027304} {\bibfield  {journal} {\bibinfo  {journal} {Phys.
  Rev. C}\ }\textbf {\bibinfo {volume} {91}},\ \bibinfo {pages} {027304}
  (\bibinfo {year} {2015})}\BibitemShut {NoStop}%
\bibitem [{\citenamefont {Agbemava}\ \emph {et~al.}(2015)\citenamefont
  {Agbemava}, \citenamefont {Afanasjev}, \citenamefont {Nakatsukasa},\ and\
  \citenamefont {Ring}}]{Agbemava2015PRC}%
  \BibitemOpen
  \bibfield  {author} {\bibinfo {author} {\bibfnamefont {S.~E.}\ \bibnamefont
  {Agbemava}}, \bibinfo {author} {\bibfnamefont {A.~V.}\ \bibnamefont
  {Afanasjev}}, \bibinfo {author} {\bibfnamefont {T.}~\bibnamefont
  {Nakatsukasa}}, \ and\ \bibinfo {author} {\bibfnamefont {P.}~\bibnamefont
  {Ring}},\ }\href {\doibase 10.1103/PhysRevC.92.054310} {\bibfield  {journal}
  {\bibinfo  {journal} {Phys. Rev. C}\ }\textbf {\bibinfo {volume} {92}},\
  \bibinfo {pages} {054310} (\bibinfo {year} {2015})}\BibitemShut {NoStop}%
\bibitem [{\citenamefont {Kucharek}\ and\ \citenamefont
  {Ring}(1991)}]{Kucharek1991ZPA}%
  \BibitemOpen
  \bibfield  {author} {\bibinfo {author} {\bibfnamefont {H.}~\bibnamefont
  {Kucharek}}\ and\ \bibinfo {author} {\bibfnamefont {P.}~\bibnamefont
  {Ring}},\ }\href {\doibase 10.1007/BF01282930} {\bibfield  {journal}
  {\bibinfo  {journal} {Z. Phys. A}\ }\textbf {\bibinfo {volume} {339}},\
  \bibinfo {pages} {23} (\bibinfo {year} {1991})}\BibitemShut {NoStop}%
\bibitem [{\citenamefont {Ring}\ and\ \citenamefont
  {Schuck}(1980)}]{Ring1980NMBP}%
  \BibitemOpen
  \bibfield  {author} {\bibinfo {author} {\bibfnamefont {P.}~\bibnamefont
  {Ring}}\ and\ \bibinfo {author} {\bibfnamefont {P.}~\bibnamefont {Schuck}},\
  }\href {https://www.springer.com/us/book/9783540212065} {\emph {\bibinfo
  {title} {{The Nuclear Many-Body Problem}}}}\ (\bibinfo  {publisher}
  {Springer-Verlag},\ \bibinfo {year} {1980})\BibitemShut {NoStop}%
\bibitem [{\citenamefont {Pan}\ \emph {et~al.}(2019)\citenamefont {Pan},
  \citenamefont {Zhang},\ and\ \citenamefont {Zhang}}]{Pan2019IJMPE}%
  \BibitemOpen
  \bibfield  {author} {\bibinfo {author} {\bibfnamefont {C.}~\bibnamefont
  {Pan}}, \bibinfo {author} {\bibfnamefont {K.}~\bibnamefont {Zhang}}, \ and\
  \bibinfo {author} {\bibfnamefont {S.}~\bibnamefont {Zhang}},\ }\href
  {\doibase https://doi.org/10.1142/S0218301319500824} {\bibfield  {journal}
  {\bibinfo  {journal} {Int. J. Mod. Phys. E}\ }\textbf {\bibinfo {volume}
  {28}},\ \bibinfo {pages} {1950082} (\bibinfo {year} {2019})}\BibitemShut
  {NoStop}%
\bibitem [{\citenamefont {In}\ \emph {et~al.}(2021)\citenamefont {In},
  \citenamefont {Papakonstantinou}, \citenamefont {Kim},\ and\ \citenamefont
  {Hong}}]{In2021IJMPE}%
  \BibitemOpen
  \bibfield  {author} {\bibinfo {author} {\bibfnamefont {E.~J.}\ \bibnamefont
  {In}}, \bibinfo {author} {\bibfnamefont {P.}~\bibnamefont
  {Papakonstantinou}}, \bibinfo {author} {\bibfnamefont {Y.}~\bibnamefont
  {Kim}}, \ and\ \bibinfo {author} {\bibfnamefont {S.-W.}\ \bibnamefont
  {Hong}},\ }\href {\doibase 10.1142/S0218301321500099} {\bibfield  {journal}
  {\bibinfo  {journal} {Int. J. Mod. Phys. E}\ }\textbf {\bibinfo {volume}
  {30}},\ \bibinfo {pages} {2150009} (\bibinfo {year} {2021})}\BibitemShut
  {NoStop}%
\bibitem [{\citenamefont {Pf\"utzner}\ \emph {et~al.}(2012)\citenamefont
  {Pf\"utzner}, \citenamefont {Karny}, \citenamefont {Grigorenko},\ and\
  \citenamefont {Riisager}}]{Pfutzner2012RMP}%
  \BibitemOpen
  \bibfield  {author} {\bibinfo {author} {\bibfnamefont {M.}~\bibnamefont
  {Pf\"utzner}}, \bibinfo {author} {\bibfnamefont {M.}~\bibnamefont {Karny}},
  \bibinfo {author} {\bibfnamefont {L.~V.}\ \bibnamefont {Grigorenko}}, \ and\
  \bibinfo {author} {\bibfnamefont {K.}~\bibnamefont {Riisager}},\ }\href
  {\doibase 10.1103/RevModPhys.84.567} {\bibfield  {journal} {\bibinfo
  {journal} {Rev. Mod. Phys.}\ }\textbf {\bibinfo {volume} {84}},\ \bibinfo
  {pages} {567} (\bibinfo {year} {2012})}\BibitemShut {NoStop}%
\bibitem [{\citenamefont {Pf\"{u}tzner}(2013)}]{Pfutzner2013PS}%
  \BibitemOpen
  \bibfield  {author} {\bibinfo {author} {\bibfnamefont {M.}~\bibnamefont
  {Pf\"{u}tzner}},\ }\href {\doibase 10.1088/0031-8949/2013/t152/014014}
  {\bibfield  {journal} {\bibinfo  {journal} {Phys. Scr.}\ }\textbf {\bibinfo
  {volume} {2013}},\ \bibinfo {pages} {014014} (\bibinfo {year}
  {2013})}\BibitemShut {NoStop}%
\bibitem [{\citenamefont {Spyrou}\ \emph {et~al.}(2012)\citenamefont {Spyrou},
  \citenamefont {Kohley}, \citenamefont {Baumann}, \citenamefont {Bazin},
  \citenamefont {Brown}, \citenamefont {Christian}, \citenamefont {DeYoung},
  \citenamefont {Finck}, \citenamefont {Frank}, \citenamefont {Lunderberg},
  \citenamefont {Mosby}, \citenamefont {Peters}, \citenamefont {Schiller},
  \citenamefont {Smith}, \citenamefont {Snyder}, \citenamefont {Strongman},
  \citenamefont {Thoennessen},\ and\ \citenamefont {Volya}}]{Spyrou2012PRL}%
  \BibitemOpen
  \bibfield  {author} {\bibinfo {author} {\bibfnamefont {A.}~\bibnamefont
  {Spyrou}}, \bibinfo {author} {\bibfnamefont {Z.}~\bibnamefont {Kohley}},
  \bibinfo {author} {\bibfnamefont {T.}~\bibnamefont {Baumann}}, \bibinfo
  {author} {\bibfnamefont {D.}~\bibnamefont {Bazin}}, \bibinfo {author}
  {\bibfnamefont {B.~A.}\ \bibnamefont {Brown}}, \bibinfo {author}
  {\bibfnamefont {G.}~\bibnamefont {Christian}}, \bibinfo {author}
  {\bibfnamefont {P.~A.}\ \bibnamefont {DeYoung}}, \bibinfo {author}
  {\bibfnamefont {J.~E.}\ \bibnamefont {Finck}}, \bibinfo {author}
  {\bibfnamefont {N.}~\bibnamefont {Frank}}, \bibinfo {author} {\bibfnamefont
  {E.}~\bibnamefont {Lunderberg}}, \bibinfo {author} {\bibfnamefont
  {S.}~\bibnamefont {Mosby}}, \bibinfo {author} {\bibfnamefont {W.~A.}\
  \bibnamefont {Peters}}, \bibinfo {author} {\bibfnamefont {A.}~\bibnamefont
  {Schiller}}, \bibinfo {author} {\bibfnamefont {J.~K.}\ \bibnamefont {Smith}},
  \bibinfo {author} {\bibfnamefont {J.}~\bibnamefont {Snyder}}, \bibinfo
  {author} {\bibfnamefont {M.~J.}\ \bibnamefont {Strongman}}, \bibinfo {author}
  {\bibfnamefont {M.}~\bibnamefont {Thoennessen}}, \ and\ \bibinfo {author}
  {\bibfnamefont {A.}~\bibnamefont {Volya}},\ }\href {\doibase
  10.1103/PhysRevLett.108.102501} {\bibfield  {journal} {\bibinfo  {journal}
  {Phys. Rev. Lett.}\ }\textbf {\bibinfo {volume} {108}},\ \bibinfo {pages}
  {102501} (\bibinfo {year} {2012})}\BibitemShut {NoStop}%
\bibitem [{\citenamefont {Kohley}\ \emph
  {et~al.}(2013{\natexlab{a}})\citenamefont {Kohley}, \citenamefont
  {Lunderberg}, \citenamefont {DeYoung}, \citenamefont {Volya}, \citenamefont
  {Baumann}, \citenamefont {Bazin}, \citenamefont {Christian}, \citenamefont
  {Cooper}, \citenamefont {Frank}, \citenamefont {Gade}, \citenamefont {Hall},
  \citenamefont {Hinnefeld}, \citenamefont {Luther}, \citenamefont {Mosby},
  \citenamefont {Peters}, \citenamefont {Smith}, \citenamefont {Snyder},
  \citenamefont {Spyrou},\ and\ \citenamefont {Thoennessen}}]{Kohley2013PRC}%
  \BibitemOpen
  \bibfield  {author} {\bibinfo {author} {\bibfnamefont {Z.}~\bibnamefont
  {Kohley}}, \bibinfo {author} {\bibfnamefont {E.}~\bibnamefont {Lunderberg}},
  \bibinfo {author} {\bibfnamefont {P.~A.}\ \bibnamefont {DeYoung}}, \bibinfo
  {author} {\bibfnamefont {A.}~\bibnamefont {Volya}}, \bibinfo {author}
  {\bibfnamefont {T.}~\bibnamefont {Baumann}}, \bibinfo {author} {\bibfnamefont
  {D.}~\bibnamefont {Bazin}}, \bibinfo {author} {\bibfnamefont
  {G.}~\bibnamefont {Christian}}, \bibinfo {author} {\bibfnamefont {N.~L.}\
  \bibnamefont {Cooper}}, \bibinfo {author} {\bibfnamefont {N.}~\bibnamefont
  {Frank}}, \bibinfo {author} {\bibfnamefont {A.}~\bibnamefont {Gade}},
  \bibinfo {author} {\bibfnamefont {C.}~\bibnamefont {Hall}}, \bibinfo {author}
  {\bibfnamefont {J.}~\bibnamefont {Hinnefeld}}, \bibinfo {author}
  {\bibfnamefont {B.}~\bibnamefont {Luther}}, \bibinfo {author} {\bibfnamefont
  {S.}~\bibnamefont {Mosby}}, \bibinfo {author} {\bibfnamefont {W.~A.}\
  \bibnamefont {Peters}}, \bibinfo {author} {\bibfnamefont {J.~K.}\
  \bibnamefont {Smith}}, \bibinfo {author} {\bibfnamefont {J.}~\bibnamefont
  {Snyder}}, \bibinfo {author} {\bibfnamefont {A.}~\bibnamefont {Spyrou}}, \
  and\ \bibinfo {author} {\bibfnamefont {M.}~\bibnamefont {Thoennessen}},\
  }\href {\doibase 10.1103/PhysRevC.87.011304} {\bibfield  {journal} {\bibinfo
  {journal} {Phys. Rev. C}\ }\textbf {\bibinfo {volume} {87}},\ \bibinfo
  {pages} {011304} (\bibinfo {year} {2013}{\natexlab{a}})}\BibitemShut
  {NoStop}%
\bibitem [{\citenamefont {Kohley}\ \emph
  {et~al.}(2013{\natexlab{b}})\citenamefont {Kohley}, \citenamefont {Baumann},
  \citenamefont {Bazin}, \citenamefont {Christian}, \citenamefont {DeYoung},
  \citenamefont {Finck}, \citenamefont {Frank}, \citenamefont {Jones},
  \citenamefont {Lunderberg}, \citenamefont {Luther}, \citenamefont {Mosby},
  \citenamefont {Nagi}, \citenamefont {Smith}, \citenamefont {Snyder},
  \citenamefont {Spyrou},\ and\ \citenamefont {Thoennessen}}]{Kohley2013PRL}%
  \BibitemOpen
  \bibfield  {author} {\bibinfo {author} {\bibfnamefont {Z.}~\bibnamefont
  {Kohley}}, \bibinfo {author} {\bibfnamefont {T.}~\bibnamefont {Baumann}},
  \bibinfo {author} {\bibfnamefont {D.}~\bibnamefont {Bazin}}, \bibinfo
  {author} {\bibfnamefont {G.}~\bibnamefont {Christian}}, \bibinfo {author}
  {\bibfnamefont {P.~A.}\ \bibnamefont {DeYoung}}, \bibinfo {author}
  {\bibfnamefont {J.~E.}\ \bibnamefont {Finck}}, \bibinfo {author}
  {\bibfnamefont {N.}~\bibnamefont {Frank}}, \bibinfo {author} {\bibfnamefont
  {M.}~\bibnamefont {Jones}}, \bibinfo {author} {\bibfnamefont
  {E.}~\bibnamefont {Lunderberg}}, \bibinfo {author} {\bibfnamefont
  {B.}~\bibnamefont {Luther}}, \bibinfo {author} {\bibfnamefont
  {S.}~\bibnamefont {Mosby}}, \bibinfo {author} {\bibfnamefont
  {T.}~\bibnamefont {Nagi}}, \bibinfo {author} {\bibfnamefont {J.~K.}\
  \bibnamefont {Smith}}, \bibinfo {author} {\bibfnamefont {J.}~\bibnamefont
  {Snyder}}, \bibinfo {author} {\bibfnamefont {A.}~\bibnamefont {Spyrou}}, \
  and\ \bibinfo {author} {\bibfnamefont {M.}~\bibnamefont {Thoennessen}},\
  }\href {\doibase 10.1103/PhysRevLett.110.152501} {\bibfield  {journal}
  {\bibinfo  {journal} {Phys. Rev. Lett.}\ }\textbf {\bibinfo {volume} {110}},\
  \bibinfo {pages} {152501} (\bibinfo {year} {2013}{\natexlab{b}})}\BibitemShut
  {NoStop}%
\bibitem [{\citenamefont {Golovkov}\ \emph {et~al.}(2004)\citenamefont
  {Golovkov}, \citenamefont {Grigorenko}, \citenamefont {Fomichev},
  \citenamefont {Oganessian}, \citenamefont {Orlov}, \citenamefont {Rodin},
  \citenamefont {Sidorchuk}, \citenamefont {Slepnev}, \citenamefont
  {Stepantsov}, \citenamefont {Ter-Akopian},\ and\ \citenamefont
  {Wolski}}]{Golovkov2004PLB}%
  \BibitemOpen
  \bibfield  {author} {\bibinfo {author} {\bibfnamefont {M.~S.}\ \bibnamefont
  {Golovkov}}, \bibinfo {author} {\bibfnamefont {L.~V.}\ \bibnamefont
  {Grigorenko}}, \bibinfo {author} {\bibfnamefont {A.~S.}\ \bibnamefont
  {Fomichev}}, \bibinfo {author} {\bibfnamefont {Y.~T.}\ \bibnamefont
  {Oganessian}}, \bibinfo {author} {\bibfnamefont {Y.~I.}\ \bibnamefont
  {Orlov}}, \bibinfo {author} {\bibfnamefont {A.~M.}\ \bibnamefont {Rodin}},
  \bibinfo {author} {\bibfnamefont {S.~I.}\ \bibnamefont {Sidorchuk}}, \bibinfo
  {author} {\bibfnamefont {R.~S.}\ \bibnamefont {Slepnev}}, \bibinfo {author}
  {\bibfnamefont {S.~V.}\ \bibnamefont {Stepantsov}}, \bibinfo {author}
  {\bibfnamefont {G.~M.}\ \bibnamefont {Ter-Akopian}}, \ and\ \bibinfo {author}
  {\bibfnamefont {R.}~\bibnamefont {Wolski}},\ }\href {\doibase
  https://doi.org/10.1016/j.physletb.2004.02.069} {\bibfield  {journal}
  {\bibinfo  {journal} {Phys. Lett. B}\ }\textbf {\bibinfo {volume} {588}},\
  \bibinfo {pages} {163} (\bibinfo {year} {2004})}\BibitemShut {NoStop}%
\bibitem [{\citenamefont {Grigorenko}\ \emph {et~al.}(2011)\citenamefont
  {Grigorenko}, \citenamefont {Mukha}, \citenamefont {Scheidenberger},\ and\
  \citenamefont {Zhukov}}]{Grigorenko2011PRC}%
  \BibitemOpen
  \bibfield  {author} {\bibinfo {author} {\bibfnamefont {L.~V.}\ \bibnamefont
  {Grigorenko}}, \bibinfo {author} {\bibfnamefont {I.~G.}\ \bibnamefont
  {Mukha}}, \bibinfo {author} {\bibfnamefont {C.}~\bibnamefont
  {Scheidenberger}}, \ and\ \bibinfo {author} {\bibfnamefont {M.~V.}\
  \bibnamefont {Zhukov}},\ }\href {\doibase 10.1103/PhysRevC.84.021303}
  {\bibfield  {journal} {\bibinfo  {journal} {Phys. Rev. C}\ }\textbf {\bibinfo
  {volume} {84}},\ \bibinfo {pages} {021303} (\bibinfo {year}
  {2011})}\BibitemShut {NoStop}%
\bibitem [{\citenamefont {Fossez}\ \emph {et~al.}(2017)\citenamefont {Fossez},
  \citenamefont {Rotureau}, \citenamefont {Michel},\ and\ \citenamefont
  {Nazarewicz}}]{Fossez2017PRC}%
  \BibitemOpen
  \bibfield  {author} {\bibinfo {author} {\bibfnamefont {K.}~\bibnamefont
  {Fossez}}, \bibinfo {author} {\bibfnamefont {J.}~\bibnamefont {Rotureau}},
  \bibinfo {author} {\bibfnamefont {N.}~\bibnamefont {Michel}}, \ and\ \bibinfo
  {author} {\bibfnamefont {W.}~\bibnamefont {Nazarewicz}},\ }\href {\doibase
  10.1103/PhysRevC.96.024308} {\bibfield  {journal} {\bibinfo  {journal} {Phys.
  Rev. C}\ }\textbf {\bibinfo {volume} {96}},\ \bibinfo {pages} {024308}
  (\bibinfo {year} {2017})}\BibitemShut {NoStop}%
\bibitem [{\citenamefont {Grigorenko}\ and\ \citenamefont
  {Zhukov}(2015)}]{Grigorenko2015PRC}%
  \BibitemOpen
  \bibfield  {author} {\bibinfo {author} {\bibfnamefont {L.~V.}\ \bibnamefont
  {Grigorenko}}\ and\ \bibinfo {author} {\bibfnamefont {M.~V.}\ \bibnamefont
  {Zhukov}},\ }\href {\doibase 10.1103/PhysRevC.91.064617} {\bibfield
  {journal} {\bibinfo  {journal} {Phys. Rev. C}\ }\textbf {\bibinfo {volume}
  {91}},\ \bibinfo {pages} {064617} (\bibinfo {year} {2015})}\BibitemShut
  {NoStop}%
\bibitem [{\citenamefont {He}\ \emph {et~al.}(2021)\citenamefont {He},
  \citenamefont {Wang}, \citenamefont {Zhang},\ and\ \citenamefont
  {Shen}}]{He2021CPC}%
  \BibitemOpen
  \bibfield  {author} {\bibinfo {author} {\bibfnamefont {X.}~\bibnamefont
  {He}}, \bibinfo {author} {\bibfnamefont {C.}~\bibnamefont {Wang}}, \bibinfo
  {author} {\bibfnamefont {K.}~\bibnamefont {Zhang}}, \ and\ \bibinfo {author}
  {\bibfnamefont {C.}~\bibnamefont {Shen}},\ }\href
  {https://doi.org/10.1088/1674-1137/ac1b99} {\bibfield  {journal} {\bibinfo
  {journal} {Chin. Phys. C}\ ,\ \bibinfo {pages} {doi:10.1088/1674}} (\bibinfo
  {year} {2021})}\BibitemShut {NoStop}%
\bibitem [{\citenamefont {Zhang}\ \emph {et~al.}(2014)\citenamefont {Zhang},
  \citenamefont {Niu}, \citenamefont {Li}, \citenamefont {Yao},\ and\
  \citenamefont {Meng}}]{Zhang2014FoP}%
  \BibitemOpen
  \bibfield  {author} {\bibinfo {author} {\bibfnamefont {Q.-S.}\ \bibnamefont
  {Zhang}}, \bibinfo {author} {\bibfnamefont {Z.-M.}\ \bibnamefont {Niu}},
  \bibinfo {author} {\bibfnamefont {Z.-P.}\ \bibnamefont {Li}}, \bibinfo
  {author} {\bibfnamefont {J.-M.}\ \bibnamefont {Yao}}, \ and\ \bibinfo
  {author} {\bibfnamefont {J.}~\bibnamefont {Meng}},\ }\href {\doibase
  10.1007/s11467-014-0413-5} {\bibfield  {journal} {\bibinfo  {journal} {Front.
  Phys.}\ }\textbf {\bibinfo {volume} {9}},\ \bibinfo {pages} {529} (\bibinfo
  {year} {2014})}\BibitemShut {NoStop}%
\bibitem [{\citenamefont {Yang}\ \emph
  {et~al.}(2021{\natexlab{b}})\citenamefont {Yang}, \citenamefont {Wang},
  \citenamefont {Zhao},\ and\ \citenamefont {Li}}]{Yang2021arXiv}%
  \BibitemOpen
  \bibfield  {author} {\bibinfo {author} {\bibfnamefont {Y.~L.}\ \bibnamefont
  {Yang}}, \bibinfo {author} {\bibfnamefont {Y.~K.}\ \bibnamefont {Wang}},
  \bibinfo {author} {\bibfnamefont {P.~W.}\ \bibnamefont {Zhao}}, \ and\
  \bibinfo {author} {\bibfnamefont {Z.~P.}\ \bibnamefont {Li}},\ }\href
  {https://arxiv.org/abs/2108.13057} {\bibfield  {journal} {\bibinfo  {journal}
  {arXiv}\ }\textbf {\bibinfo {volume} {2108:13057}} (\bibinfo {year}
  {2021}{\natexlab{b}})}\BibitemShut {NoStop}%
\bibitem [{\citenamefont {Dong}\ \emph {et~al.}(2018)\citenamefont {Dong},
  \citenamefont {Zhang}, \citenamefont {Zuo}, \citenamefont {Gu}, \citenamefont
  {Wang},\ and\ \citenamefont {Sun}}]{Dong2018PRC}%
  \BibitemOpen
  \bibfield  {author} {\bibinfo {author} {\bibfnamefont {J.~M.}\ \bibnamefont
  {Dong}}, \bibinfo {author} {\bibfnamefont {Y.~H.}\ \bibnamefont {Zhang}},
  \bibinfo {author} {\bibfnamefont {W.}~\bibnamefont {Zuo}}, \bibinfo {author}
  {\bibfnamefont {J.~Z.}\ \bibnamefont {Gu}}, \bibinfo {author} {\bibfnamefont
  {L.~J.}\ \bibnamefont {Wang}}, \ and\ \bibinfo {author} {\bibfnamefont
  {Y.}~\bibnamefont {Sun}},\ }\href {\doibase 10.1103/PhysRevC.97.021301}
  {\bibfield  {journal} {\bibinfo  {journal} {Phys. Rev. C}\ }\textbf {\bibinfo
  {volume} {97}},\ \bibinfo {pages} {021301} (\bibinfo {year}
  {2018})}\BibitemShut {NoStop}%
\bibitem [{\citenamefont {Dong}\ \emph {et~al.}(2019)\citenamefont {Dong},
  \citenamefont {Shang}, \citenamefont {Zuo}, \citenamefont {Niu},\ and\
  \citenamefont {Sun}}]{Dong2019NPA}%
  \BibitemOpen
  \bibfield  {author} {\bibinfo {author} {\bibfnamefont {J.~M.}\ \bibnamefont
  {Dong}}, \bibinfo {author} {\bibfnamefont {X.~L.}\ \bibnamefont {Shang}},
  \bibinfo {author} {\bibfnamefont {W.}~\bibnamefont {Zuo}}, \bibinfo {author}
  {\bibfnamefont {Y.~F.}\ \bibnamefont {Niu}}, \ and\ \bibinfo {author}
  {\bibfnamefont {Y.}~\bibnamefont {Sun}},\ }\href {\doibase
  https://doi.org/10.1016/j.nuclphysa.2019.01.003} {\bibfield  {journal}
  {\bibinfo  {journal} {Nucl. Phys. A}\ }\textbf {\bibinfo {volume} {983}},\
  \bibinfo {pages} {133} (\bibinfo {year} {2019})}\BibitemShut {NoStop}%
\bibitem [{\citenamefont {Dong}\ and\ \citenamefont
  {Shang}(2020)}]{Dong2020PRC}%
  \BibitemOpen
  \bibfield  {author} {\bibinfo {author} {\bibfnamefont {J.~M.}\ \bibnamefont
  {Dong}}\ and\ \bibinfo {author} {\bibfnamefont {X.~L.}\ \bibnamefont
  {Shang}},\ }\href {\doibase 10.1103/PhysRevC.101.014305} {\bibfield
  {journal} {\bibinfo  {journal} {Phys. Rev. C}\ }\textbf {\bibinfo {volume}
  {101}},\ \bibinfo {pages} {014305} (\bibinfo {year} {2020})}\BibitemShut
  {NoStop}%
\end{thebibliography}

%

\end{document}